\documentclass[prc,superscriptaddress,showpacs,psfig,showkeys,nofootinbib]{revtex4}
\usepackage{amsmath}
\usepackage{amssymb}
\usepackage{amsfonts}
\usepackage{graphicx}
\usepackage{epsfig}
\usepackage{dcolumn}
\usepackage{hyperref}
\usepackage{epstopdf}
\usepackage{color}
\usepackage[utf8]{inputenc}
\usepackage{amsthm}
\usepackage{fontenc}
\usepackage{bm}
\RequirePackage{slashed}\usepackage{cancel}
\usepackage[normalem]{ulem}

\usepackage{array,booktabs}
\newcolumntype{C}{>{$\displaystyle}c<{$}}
\usepackage{soul}
\usepackage{color,soul}

\usepackage{ulem}

\newcommand \Pomeron {I\!\!P}

\begin{document}


\title{Nuclear suppression of coherent $J/\psi$ photoproduction in heavy-ion UPCs and leading twist nuclear shadowing}

\author{V. Guzey}

\affiliation{University of Jyvaskyla, Department of Physics, P.O. Box 35, FI-40014 University
of Jyvaskyla, Finland and Helsinki Institute of Physics, P.O. Box 64, FI-00014
University of Helsinki, Finland}

\author{M. Strikman}

\affiliation{Pennsylvania  State  University,  University  Park,  PA,  16802,  USA}

\date{\today}

\begin{abstract}

We determine the nuclear suppression factor $S_{Pb}(x)$, where $x=M_{J/\psi}^2/W_{\gamma p}^2$ with $M_{J/\psi}$ the $J/\psi$ mass and
$W_{\gamma p}$ the photon-nucleon energy,
for the cross section of coherent $J/\psi$ photoproduction in heavy-ion 
ultraperipheral collisions (UPCs) at the Large Hadron Collider (LHC) and Relativistic Heavy Ion Collider (RHIC)
by performing the $\chi^2$ fit to all available data on the cross section $d\sigma^{AA \to J/\psi AA}/dy$ 
as a function of the $J/\psi$ rapidity $y$ and the photoproduction cross section $\sigma^{\gamma A \to J/\psi A}(W_{\gamma p})$ as a function of $W_{\gamma p}$. We find that while the $d\sigma^{AA \to J/\psi AA}/dy$ data alone constrain $S_{Pb}(x)$ for $x \geq 10^{-3}$,
the combined $d\sigma^{AA \to J/\psi AA}/dy$  and $\sigma^{\gamma A \to J/\psi A}(W_{\gamma p})$ data allow us to determine
$S_{Pb}(x)$ in the wide interval $10^{-5} < x < 0.05$. In particular, the data favor $S_{Pb}(x)$, which decreases with a decrease of $x$ in the $10^{-4} < x < 0.01$ interval, and can be both decreasing or constant for $x< 10^{-4}$.
Identifying $S_{Pb}(x)$ with the ratio of the gluon distributions in Pb and the proton $R_g(x,Q_0^2)=g_A(x,Q_0^2)/[A g_p(x,Q_0^2)]$, 
we demonstrate that the leading twist approximation (LTA) for nuclear shadowing provides a good description of all the data on 
$d\sigma^{AA \to J/\psi AA}/dy$ and $\sigma^{\gamma A \to J/\psi A}(W_{\gamma p})$ as well as on the experimental values for $S_{Pb}(x)$ derived from
 $\sigma^{\gamma A \to J/\psi A}(W_{\gamma p})$. We also show that modern nuclear PDFs reasonably reproduce $S_{Pb}(x)$ as well.
 
\end{abstract}

\maketitle

\section{Introduction}
\label{sec:intro}

Ultraperipheral collisions (UPCs) of heavy ions have emerged as a novel and powerful tool to access the partonic structure of nuclei and the dynamics of strong interactions at high energies in quantum chromodynamics (QCD)~\cite{Bertulani:2005ru,Baltz:2007kq}. In UPCs, colliding ions pass each other at large distances in the transverse plane, and their interactions are mediated by quasi-real photons produced by these ions. It effectively turns heavy-ion UPCs at the Large Hadron Collider (LHC) and the Relativistic Heavy Ion Collider (RHIC) into a high-energy and high-intensity photon-nucleus collider. Hence, it gives a unique opportunity to study small-$x$ QCD in nuclei using hard photon-nucleus scattering in the previously unaccessible kinematics, which is complimentary 
to that of the planned Electron-Ion Collider~\cite{Accardi:2012qut,AbdulKhalek:2021gbh}.

Since pioneering experimental results on heavy-ion UPCs at RHIC more than 20 years ago, the main focus of UPC studies has been photoproduction of light
and heavy vector mesons. In particular, it has been firmly established that the cross section of coherent $J/\psi$ photoproduction in Pb-Pb UPCs  
at the LHC and Au-Au UPCs at RHIC is suppressed compared to the impulse approximation expectations. The origin of this suppression is one of open 
questions of high-energy QCD: the successful predictions of the leading twist approach (LTA) for gluon nuclear shadowing~\cite{Guzey:2013xba,Guzey:2013qza} compete with the description within the color dipole model including the non-linear effect of gluon saturation~\cite{Bendova:2020hbb,Cepila:2017nef,Mantysaari:2017dwh,Mantysaari:2022sux}.

A first model-independent extraction of the nuclear suppression factor $S_{Pb}(x)$ for coherent $J/\psi$ photoproduction in Pb-Pb UPCs at the LHC has 
been carried out in~\cite{Guzey:2020ntc}. Here $x=M_{J/\psi}^2/W_{\gamma p}^2$, where $M_{J/\psi}$ is the $J/\psi$ mass and $W_{\gamma p}$ is the photon-nucleon center-of-mass energy, is the longitudinal momentum fraction, which can be associated with that of the probed nuclear parton distributions (PDFs). 
However, since then, there have appeared new data on coherent $J/\psi$ photoproduction in Pb-Pb UPCs 
accompanied by electromagnetic excitation of colliding ions with forward neutron emission at the LHC~\cite{CMS:2023snh,ALICE:2023jgu}, which 
are sensitive to $S_{Pb}(x)$ in the previously unexplored region of small $x$ down to $x \approx 10^{-5}$. In addition, the new STAR measurement at RHIC~\cite{STAR:2023gpk} has supplied an important reference points at intermediate $x =0.015$.

In this paper, we perform a new analysis of $S_{Pb}(x)$, where it is determined using the $\chi^2$ fit to all data on coherent $J/\psi$ photoproduction in heavy-ion UPCs available to date. 
We find that while the data on the rapidity-differential cross section $d\sigma^{AA \to J/\psi AA}/dy$ do not constrain $S_{Pb}(x)$ for $x < 10^{-3}$,
their combination with the data on the nuclear photoproduction cross section $\sigma^{\gamma A \to J/\psi A}(W_{\gamma p})$ allows us to determine
$S_{Pb}(x)$ in the wide interval $10^{-5} < x < 0.05$. The data favor $S_{Pb}(x)$, which decreases with a decrease of $x$ in the $10^{-4} < x < 0.01$ 
interval. For $ x < 10^{-4}$, both constant and decreasing $S_{Pb}(x)$ can be accommodated by the data.

Identifying $S_{Pb}(x)$ with the ratio of the gluon distributions in a heavy nucleus and the proton $R_g(x,Q_0^2)=g_A(x,Q_0^2)/[A g_p(x,Q_0^2)]$, 
where $Q_0^2=3$ GeV$^2$ is set by the charm quark mass,
we find that
the LTA predictions provide a good description of all the data on $d\sigma^{AA \to J/\psi AA}/dy$ and $\sigma^{\gamma A \to J/\psi A}(W_{\gamma p})$ as well on the experimental values of $S_{Pb}(x)$ derived from the $\sigma^{\gamma A \to J/\psi A}(W_{\gamma p})$ photoproduction cross section.
This serves as another strong argument supporting the presence of large leading twist gluon nuclear shadowing at small $x$. We argue that the description of $S_{Pb}(x)$ at intermediate $x \approx 10^{-2}$ is also fair and can be further improved by modeling of the gluon antishadowing. We also show that within large uncertainties of modern state-of-the-art nuclear PDFs, the EPPS21, nCTEQ15HQ, and nNNPDF3.0 nuclear PDFs give a reasonable description of $S_{Pb}(x)$
in the entire range of $x$.

The rest of this paper is organized as follows. In Sec.~\ref{sec:suppression}, we outline main ingredients for the calculation of the $d\sigma^{AA \to J/\psi AA}/dy$ and $\sigma^{\gamma A \to J/\psi A}(W_{\gamma p})$ cross sections and the procedure for a model-independent extraction of $S_{Pb}(x)$ using the $\chi^2$ fit to the UPC data on these cross sections. We present and discuss results of the fit and compare predictions of the fit solutions with
$d\sigma^{AA \to J/\psi AA}/dy$ and $\sigma^{\gamma A \to J/\psi A}(W_{\gamma p})$ used in the fit and with the experimental values of $S_{Pb}(x)$.
In Sec.~\ref{sec:lta}, we recapitulate the LTA results for $R_g(x,Q_0^2)$ and compare the LTA predictions
with the data on $d\sigma^{AA \to J/\psi AA}/dy$, $\sigma^{\gamma A \to J/\psi A}(W_{\gamma p})$ and $S_{Pb}(x)$. 
The latter observable is also compared with predictions of the EPPS21, nCTEQ15HQ, and nNNPDF3.0 nuclear PDFs.
We demonstrate that modifications of LTA, which include dynamical modeling of gluon antishadowing and impact parameter dependent nuclear shadowing, can further improve the description of the UPC data at both intermediate and small $x$. 
Our conclusions are
given in Sec.~\ref{sec:conclusions}.

\section{Nuclear suppression factor for cross section of coherent $J/\psi$ photoproduction in UPCs}
\label{sec:suppression}

\subsection{UPC cross section, photon flux, and definition of the suppression factor}
\label{subsec:def}

In the equivalent photon approximation~\cite{Budnev:1975poe}, the cross section of coherent $J/\psi$ photoproduction in heavy-ion UPCs integrated over the
$J/\psi$ transverse momentum has the following form~\cite{Bertulani:2005ru,Baltz:2007kq}
\begin{equation}
\frac{d\sigma^{AA \to J/\psi AA}(y)}{dy}=\left[k\frac{dN_{\gamma/A}}{dk} \sigma^{\gamma A \to J/\psi A}(W_{\gamma p})  \right]_{k=k^{+}}+\left[k\frac{dN_{\gamma/A}}{dk} \sigma^{\gamma A \to J/\psi A}(W_{\gamma p})\right]_{k=k^{-}} \,,
\label{eq:upc_cs}
\end{equation}
where $k dN_{\gamma/A}/dk$ is the photon flux and $\sigma^{\gamma A \to J/\psi A}(W_{\gamma p})$ is the nuclear photoproduction cross section.
The UPC cross section depends on the $J/\psi$ rapidity $y$, which determines the photon momentum (energy) in the center-of-mass frame $k^{\pm}=(M_{J/\psi}/2) e^{\pm y}$, where $M_{J/\psi}$ is the $J/\psi$ mass.
The underlying photoproduction cross section is a function of the center-of-mass photon-nucleon energy
$W_{\gamma p}=\sqrt{ 4 k E_A}=\sqrt{ 4 k \gamma_L m_N}=\sqrt{ 2 \gamma_L m_N M_{J/\psi} e^{\pm y}}$, where
$E_A$ is the nuclear beam energy, $\gamma_L$ is the corresponding Lorentz factor, and $m_N$ is the nucleon mass.
The two-fold ambiguity in the photon momentum $k^{\pm}$ and the photon-nucleon energy $W_{\gamma p}$ at given $y \neq 0$ is a consequence of the possibility for both colliding ions to serve as a source of the photons and as a target, which is reflected in the presence of two terms in Eq.~(\ref{eq:upc_cs}).

In principle, the UPC cross section also contains the interference contribution of the two terms in Eq~(\ref{eq:upc_cs}). However, since 
it is concentrated at very small values of the $J/\psi$ transverse momentum $p_T$, $p_T < 10$ MeV/c for RHIC and
$p_T < 4$ MeV/c for the LHC~\cite{Klein:1999gv}, its contribution to the $p_T$-integrated UPC cross section is vanishingly small and has been neglected in Eq~(\ref{eq:upc_cs}).

The flux of equivalent photons is obtained by combining elements of quantum electrodynamics (QED) and soft strong hadron-hadron interactions and
is given by the following convolution over the impact parameter $\vec{b}$,
\begin{equation}
k \frac{dN_{\gamma/A}(k)}{dk} = \int d^2 \vec{b}\, N_{\gamma/A}(k, \vec{b}) \Gamma_{AA}(\vec{b}) \,.
\label{eq:flux}
\end{equation}
In this equation, $N_{\gamma/A}(k, \vec{b})$ is the photon flux produced by an ultrarelativistic nucleus at the transverse distance $\vec{b}$ from its center~\cite{Vidovic:1992ik},
\begin{equation}
N_{\gamma}(k, \vec{b})=\frac{Z^2 \alpha_{\rm e.m.}}{\pi^2}\left|\int_0^{\infty} dk_{\perp} \frac{k_{\perp}^2 F_{ch}(k_{\perp}^2+\omega^2/\gamma_L^2)}{k_{\perp}^2+\omega^2/\gamma_L^2} 
J_1(k_{\perp} |\vec{b}|)\right|_{\omega=k}^2 \,,
\label{eq:flux_b}
\end{equation} 
where $\alpha_{\rm e.m.}$ is the fine-structure constant, $Z$ is the nucleus electric charge, $F_{ch}(t)$ is the nucleus charge form factor normalized to unity,
$F_{ch}(t=0)=1$,
$k_{\perp}$ is the photon transverse momentum, 
and $J_1$ is the Bessel function of the first kind.

The factor $\Gamma_{AA}(\vec{b})$ represents the probability of the absence of strong inelastic nucleus-nucleus interactions at the transverse distance $\vec{b}$
between the centers of the colliding nuclei, which can be evaluated using the optical limit of the Glauber model,
\begin{equation}
\Gamma_{AA}(\vec{b}) = \exp\left(-\sigma_{NN}(s_{NN}) \int d^2 \vec{b^{\prime}}\, T_A(\vec{b^{\prime}}) T_A(\vec{b}-\vec{b^{\prime}})\right) \,,
\label{eq:Gamma_AA}
\end{equation}
where $\sigma_{NN}(s_{NN})$ is the total nucleon-nucleon cross section~\cite{ParticleDataGroup:2022pth}, 
and $T_A(\vec{b})=\int dz \rho_A(\vec{b},z)$ is the nuclear thickness function (optical density) with $\rho_A(\vec{r})$ the nuclear density~\cite{DeVries:1987atn}.
The use of $\Gamma_{AA}(\vec{b})$, which strongly suppresses the contribution of $|\vec{b}| < 2 R_A$ in Eq.~(\ref{eq:flux}), where $R_A$ is the nucleus radius,
allows one to avoid using an artificial and model-dependent cutoff on the relevant range of impact parameters in the calculation of the photon flux, see the discussion in~\cite{Nystrand:2004vn}.
Note that this derivation implies that $|\vec{b}| \gg 2 R_A$, which simplifies the impact parameter dependence on the nuclear target side; for corrections beyond this approximation, see~\cite{Eskola:2024fhf}.

Information on the small-$x$ QCD dynamics and nuclear structure is encoded in the underlying cross section 
$\sigma^{\gamma A \to J/\psi A}(W_{\gamma p})$. 
It is convenient to quantify its nuclear modifications with respect to the impulse approximation (IA) by introducing the nuclear suppression factor $S_{Pb}(x)$~\cite{Guzey:2013xba,Guzey:2013qza}
\begin{equation}
\sigma^{\gamma A \to J/\psi A}(W_{\gamma p})=[S_{Pb}(x)]^2 \sigma^{\gamma A \to J/\psi A}_{\rm IA}(W_{\gamma p}) \,,
\label{eq:R}
\end{equation}
where $x=M_{J/\psi}^2/W_{\gamma p}^2$. Note that we use the subscript ``Pb'' because apart from the $^{197}$Au STAR data point,
the rest of the data used in our analysis comes from LHC measurements using the nucleus of $^{208}$Pb as a a target.   
The IA expression for the cross section of coherent $J/\psi$ photoproduction on a nucleus integrated over the momentum transfer $t \approx -p_T^2$ is 
\begin{equation}
\sigma^{\gamma A \to J/\psi A}_{\rm IA}(W_{\gamma p})=\frac{d\sigma^{\gamma +p \to J/\psi+ p}(W_{\gamma p}, t=0)}{dt} \int_{|t_{\rm min}|}^{\infty} dt |F_A(t)|^2 \,,
\label{eq:cs_IA}
\end{equation}
where $d\sigma^{\gamma p \to J/\psi p}(t=0)/dt$ is the differential cross section on the proton at $t=0$, $F_A(t)$ is the nuclear form factor normalized to the number of nucleons $A$, $F_A(t=0)=A$,  
and $|t_{\rm min}|=x^2 m_N^2$ is the minimal momentum transfer squared.
The IA cross section in Eq.~(\ref{eq:cs_IA}) takes into account nuclear coherence, but neglects all other nuclear effects. 
It can be calculated in a model-independent way using reliable data-based parametrizations of $F_A(t)$ and
$d\sigma^{\gamma +p \to J/\psi+ p}(W_{\gamma p}, t=0)/dt$, see Sec.~\ref{subsec:proton}.

In terms of $S_{Pb}(x)$, the UPC cross section in Eq.~(\ref{eq:upc_cs}) can be re-written in the following convenient form,
\begin{equation}
\frac{d\sigma^{AA \to J/\psi AA}(y)}{dy}=\left[k\frac{dN_{\gamma}}{dk} \sigma^{\gamma A \to J/\psi A}_{\rm IA}(W_{\gamma p}) (S_{Pb}(x))^2 \right]_{k=k^{+}}+\left[k\frac{dN_{\gamma}}{dk} \sigma^{\gamma A \to J/\psi A}_{\rm IA}(W_{\gamma p})(S_{Pb}(x))^2\right]_{k=k^{-}} \,.
\label{eq:upc_cs_2}
\end{equation}
The main advantage of using $S_{Pb}(x)$ is that it allows one to disentangle the ambiguous relation between $y$ and $x$, see the discussion above. As a result, $S_{Pb}(x)$ can be determined by fitting to UPC data on $d\sigma^{AA \to J/\psi AA}/dy$ at different $\sqrt{s_{NN}}$, 
i.e., one can combine the Run 1 and Run 2 data in the same fit
and also add other observables such as the recently measured $\sigma^{\gamma A \to J/\psi A}(W_{\gamma p})$ nuclear photoproduction
cross section.

\subsection{Cross section of exclusive $J/\psi$ photoproduction on the proton}
\label{subsec:proton}

The expression for the cross section of coherent $J/\psi$ photoproduction on a nucleus in IA requires as input
the $\gamma+p \to J/\psi+p$ differential cross section on the proton at $t=0$. 
In our analysis, we use two different parametrizations of this cross section as a function of $W_{\gamma p}$.

The first one is based on the 2013 H1 fit for the $t$-integrated cross section of elastic $J/\psi$ photoproduction
on the proton~\cite{H1:2013okq}
\begin{equation}
\sigma^{\gamma+p \to J/\psi+p}(W_{\gamma p})_{|\rm H1}=N_{\rm el} (W_{\gamma p}/W_0)^{\delta_{\rm el}} \,,
\label{eq:fit_H1}
\end{equation}
where $N_{\rm el}=81 \pm 3$ nb, $\delta_{\rm el}=0.67 \pm 0.03$, and $W_0=90$ GeV.
It is converted into the differential cross section at $t=0$ by assuming an exponential and energy-dependent slope $B(W_{\gamma p})$ of the $t$-dependence of $d\sigma^{\gamma+p \to J/\psi+p}/dt$,
\begin{equation}
\frac{d\sigma^{\gamma+p \to J/\psi+p}(W_{\gamma p},t=0)}{dt}_{|\rm H1}=B(W_{\gamma p}) \sigma^{\gamma+p \to J/\psi+p}(W_{\gamma p})_{|\rm H1} \,.
\label{eq:fit_H1_t0}
\end{equation}
For $B(W_{\gamma p})$, we use the Regge-motivated parametrization
\begin{equation}
B(W_{\gamma p})=B_0 + 4 \alpha_{\Pomeron}^{\prime} \ln(W_{\gamma p}/W_0) \,,
\label{eq:Bslope}
\end{equation}
where the parameters are taken from the 2002 ZEUS analysis~\cite{ZEUS:2002wfj}, 
\begin{eqnarray}
B_0 &=& 4.15 \pm 0.05 (stat.)^{+0.30}_{-0.18}(syst.) \ {\rm GeV}^{-2}\,, \nonumber\\
\alpha_{\Pomeron}^{\prime}& =& 0.116 \pm 0.026(stat.)^{+0.10}_{-0.025}(syst.) \ {\rm GeV}^{-2} \,.
\label{eq:Bslop2}
\end{eqnarray}

The second parametrization directly fits the $d\sigma^{\gamma+p \to J/\psi+p}/dt$ differential cross section at $t=0$~\cite{Guzey:2013xba}
\begin{equation}
\frac{d\sigma^{\gamma+p \to J/\psi+p}(W_{\gamma p},t=0)}{dt}_{|\rm GKSZ}=C_0 \left[1-\frac{(M_{J/\psi}+m_N)^2}{W_{\gamma p}^2} \right]^{1.5} (W_{\gamma p}^2/W_0^2)^{\delta} \,,
\label{eq:fit_K}
\end{equation}
where $C_0=342 \pm 8$ nb/GeV$^2$, $\delta = 0.40 \pm 0.01$, and $W_0=100$ GeV.
The cross section integrated over $t$ is obtained by combining Eq.~(\ref{eq:fit_K}) with the slope $B(W_{\gamma p})$  of Eq.~(\ref{eq:Bslope}),
\begin{equation}
\sigma^{\gamma+p \to J/\psi+p}(W_{\gamma p})_{|\rm GKSZ}=\frac{1}{B(W_{\gamma p})} \frac{d\sigma^{\gamma+p \to J/\psi+p}(W_{\gamma p},t=0)}{dt}_{|\rm GKSZ} \,.
\label{eq:fit_K_2}
\end{equation}

\begin{figure}[t]
\centerline{%
\includegraphics[width=9cm]{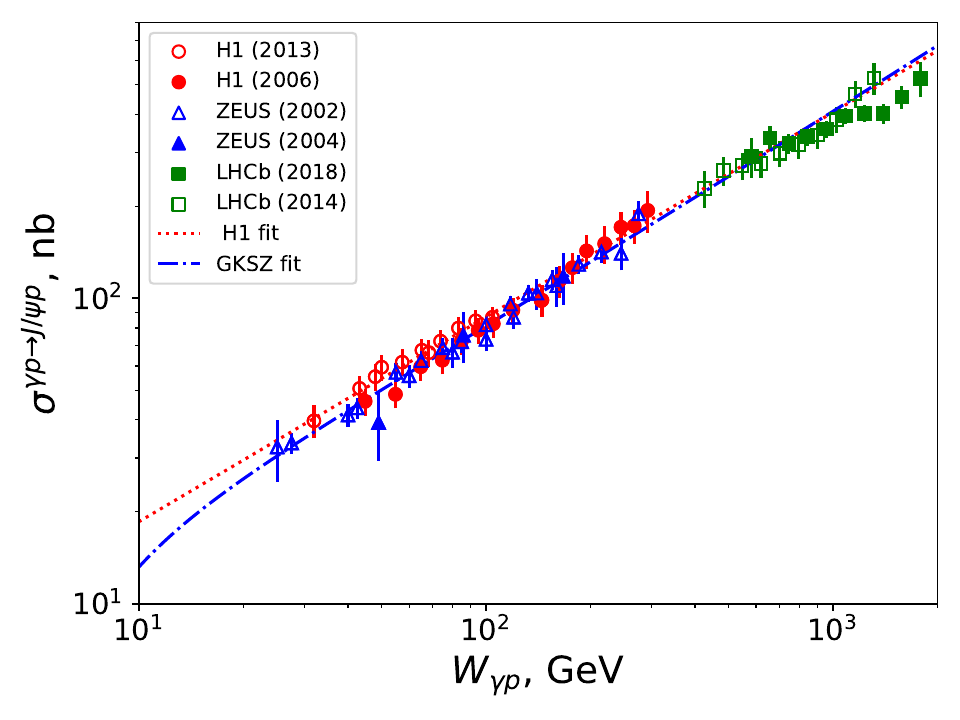}
\includegraphics[width=9cm]{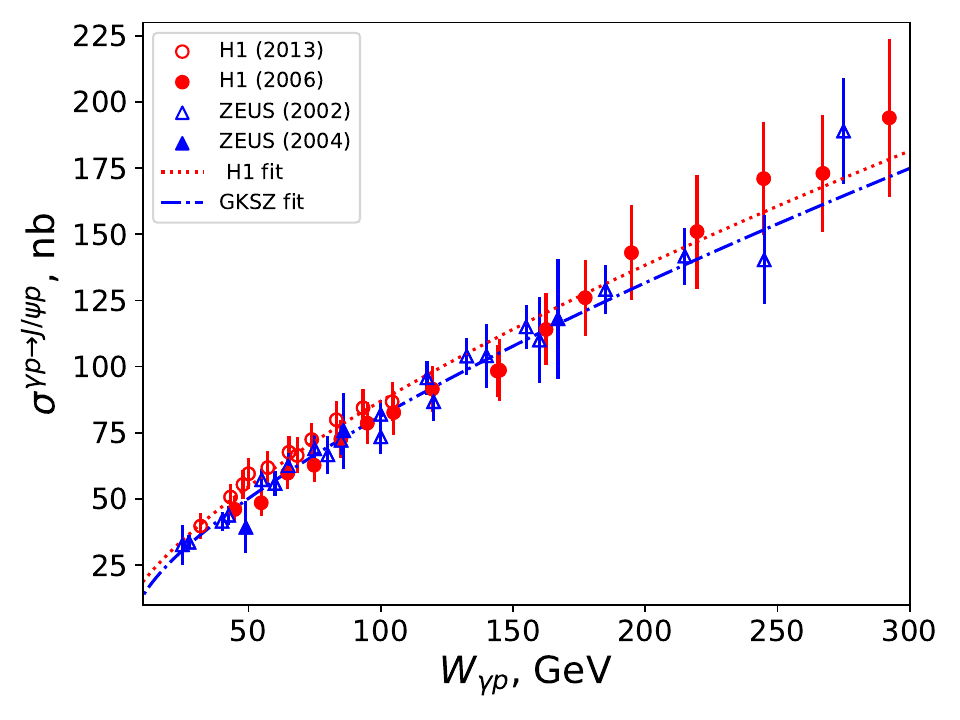}
}
\caption{The $t$-integrated $\sigma^{\gamma+p \to J/\psi+p}(W_{\gamma p})$ cross section as a function of $W_{\gamma p}$: two parametrizations (see text for details) and the HERA and LHC data. The right panel emphasizes the $W_{\gamma p}$ range covered by the HERA data.}
\label{fig:gp}
\end{figure}

A comparison of results of these two parametrizations with the data on the $t$-integrated $\sigma^{\gamma+p \to J/\psi+p}(W_{\gamma p})$ cross section 
as a function of $W_{\gamma p}$
is shown in Fig.~\ref{fig:gp}.
The data include the 2013~\cite{H1:2013okq} and 2006~\cite{H1:2005dtp} H1 data 
(we do not include the 2000 H1 data~\cite{H1:2000kis}, which is superseded by the newer H1 data), 
the 2002~\cite{ZEUS:2002wfj} and 2004~\cite{ZEUS:2004yeh} ZEUS data, and the 2014~\cite{LHCb:2014acg} and 2018~\cite{LHCb:2018rcm} LHCb data.
The left panel displays the results on a logarithmic $W_{\gamma p}$ scale, while the right panel does it on a linear $W_{\gamma p}$ scale
to emphasize the $W_{\gamma p}$ range covered by the HERA data, which is most relevant for Pb-Pb UPCs at the LHC. 
One can see from the figure that both fits describe the data very well in the entire range of $W_{\gamma p}$.

Note that while both parametrizations of $d\sigma^{\gamma+p \to J/\psi+p}(t=0)/dt$ are suitable for the calculation of 
$\sigma^{\gamma A \to J/\psi A}_{\rm IA}(W_{\gamma p})$ and $S_{Pb}(x)$, their difference somewhat affects the values of $S_{Pb}(x)$
at large $x$, see the discussion in Sec.~\ref{subsec:fit} and \ref{subsec:fit_comp}.

\subsection{Fitting the nuclear suppression factor $S_{Pb}(x)$}
\label{subsec:fit}

Before attempting to describe and interpret the data on coherent $J/\psi$ photoproduction in heavy-ion UPCs at the LHC and RHIC using theoretical models, it is important to determine the nuclear suppression factor $S_{Pb}(x)$ as model-independently as possible directly from the data, see Ref.~\cite{Guzey:2020ntc}.

The available data sample consists of 20 data points on the $d\sigma^{AA \to J/\psi AA}/dy$ Pb-Pb UPC cross section as a function of the $J/\psi$ rapidity $y$, one STAR data point on $d\sigma^{AA \to J/\psi AA}/dy$ in Au-Au UPCs at RHIC,  
and 15 points on the $\sigma^{\gamma A \to J/\psi A}(W_{\gamma p})$ photoproduction cross section on Pb as function of $W_{\gamma p}$. The $d\sigma^{AA \to J/\psi AA}/dy$ data include: 
\begin{itemize}
\item 
3 points from Run 1 at $\sqrt{s_{NN}}=2.76$ TeV:  ALICE data points at mid-rapidity~\cite{ALICE:2013wjo} and in the rapidity interval $3.6 < |y| < 2.6$~\cite{ALICE:2012yye} and the CMS point in the  $1.8 < |y| < 2.3$ interval~\cite{CMS:2016itn}
\item 
14 points from  Run 2 at $\sqrt{s_{NN}}=5.02$ TeV: ALICE data points (averaged over the $\mu^{+} \mu^{-}$ and $e^{+} e^{-}$ decay channels) at midrapidity~\cite{ALICE:2021gpt} and ALICE~\cite{ALICE:2019tqa} and LHCb~\cite{LHCb:2022ahs} points at 
forward rapidity (we do not use the earlier 2015 LHCb data~\cite{LHCb:2021bfl}, which are superseded 
by the 2018 LHCb data)
\item 3 CMS data points at intermediate rapidity  at $\sqrt{s_{NN}}=5.02$ TeV (Run 2) without neutron multiplicity selection, which corresponds to a sum of all neutron channels AnAn~\cite{CMS:2023snh}
\item One STAR data point at central rapidities $|y|< 1$ at $\sqrt{s_{NN}}=200$ GeV~\cite{STAR:2023gpk}.
\end{itemize}

The data on $\sigma^{\gamma A \to J/\psi A}(W_{\gamma p})$ come from measurements of coherent $J/\psi$ photoproduction
in Pb-Pb UPCs accompanied by electromagnetic excitation of colliding ions with forward neutron emission during Run 2 at the LHC with $\sqrt{s_{NN}}=5.02$ TeV and include 6 points by CMS~\cite{CMS:2023snh} and 9 points by ALICE~\cite{ALICE:2023jgu}.
Measurements in different classes of forward-neutron multiplicities, which
are determined with energy deposits in zero degree calorimeters (ZDCs), allowed one to disentangle the two-fold ambiguity of the UPC cross section 
in Eq.~(\ref{eq:upc_cs}) and to essentially model-independently extract the photoproduction cross section, see the discussion of the method in~\cite{Guzey:2013jaa}.

\begin{figure}[t]
\centerline{%
\includegraphics[width=10cm]{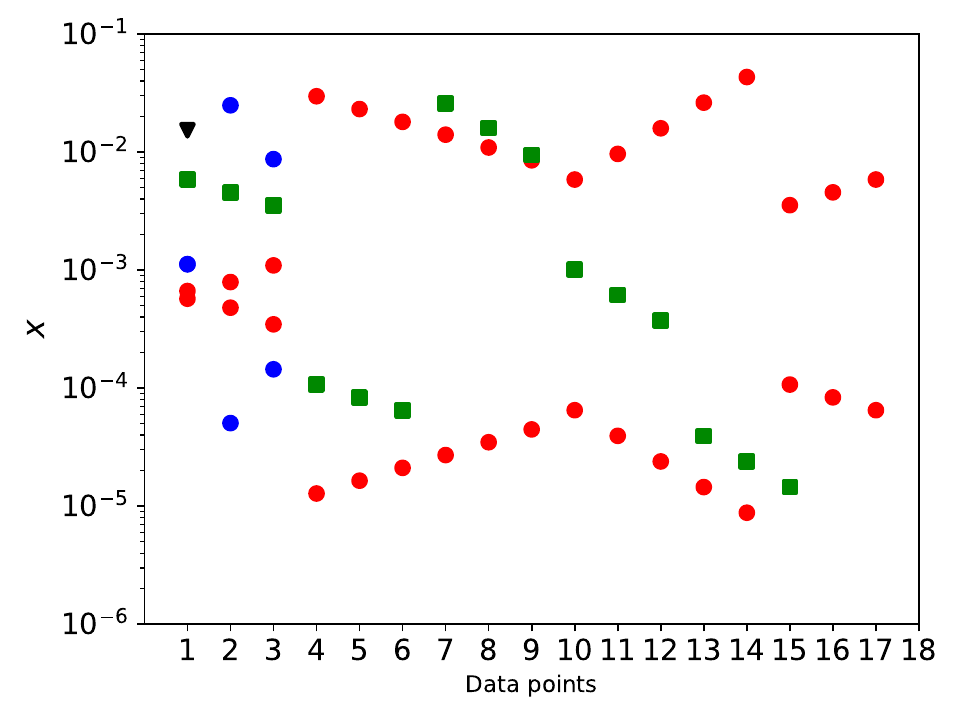}
}
\caption{The values of $x=M_{J/\psi}^2/W_{\gamma p}^2$ probed by the UPC data on $d\sigma^{AA \to J/\psi AA}/dy$ (blue and red circles for Runs 1 and 2 at the LHC, respectively, and the black triangle for RHIC)
and the photoproduction cross section $\sigma^{\gamma A \to J/\psi A}(W_{\gamma p})$ (green squares).}
\label{fig:x_coverage}
\end{figure}

Figure~\ref{fig:x_coverage} shows the values of $x=M_{J/\psi}^2/W_{\gamma p}^2$ probed by the available data on coherent $J/\psi$ photoproduction in 
heavy-ion UPCs.
The red and blue points correspond to $d\sigma^{AA \to J/\psi AA}/dy$ at Run 2 and Run 1 at the LHC, respectively, 
while the green squares correspond to the data on $\sigma^{\gamma A \to J/\psi A}(W_{\gamma p})$ at the LHC.
Note that the red and blue points are doubled (``mirrored'') because in Eq.~(\ref{eq:upc_cs_2}), each value of the $J/\psi$ rapidity 
$y \neq 0$ corresponds to two values of $x$.
The black triangle denotes the average $\langle x \rangle$ value probed by STAR, which overlaps with the LHC measurements. 
One can see from the figure that the data span a wide range in $x$ covering $10^{-5} < x < 0.05$.

One should keep in mind that for sufficiently large values of $|y|$, the UPC cross section in Eq.~(\ref{eq:upc_cs}) is dominated by the contribution
of low-energy photons with $k=(M_{J/\psi}/2) e^{-|y|}$ corresponding to the values of $W_{\gamma p}$ smaller than those covered by the HERA data in Fig.~\ref{fig:gp}.
Taking $W_{\gamma p, \rm min}=25$ GeV, we find that the rapidity ranges $|y|_{\rm max} > 2.6$ for Run 1 at the LHC
and $|y|_{\rm max} > 3.2$ for Run 2 correspond to $W_{\gamma p} < W_{\gamma p, \rm min}$, where one has to rely on the small-$W_{\gamma p}$
extrapolation of the $d\sigma^{\gamma+p \to J/\psi+p}(W_{\gamma p},t=0)/dt$ parametrizations discussed in Sec.~\ref{subsec:proton}.

Therefore, it is natural to model the fitting function for $S_{Pb}(x)$ as a function of $\ln x$.
Dividing the $10^{-5} < x < 0.1$ range into 4 intervals $x_i \leq x \leq x_{i-1}$, where $x_i=10^{-i}$ and $2 \leq i \leq 5$, 
we assume the following piece-wise form for $S_{Pb}(x)$ on each of the intervals, 
\begin{equation}
S_{Pb}(x;y_i)_{|\rm{Fit\, 1}}= y_i +\frac{y_{i-1}-y_i}{\ln(x_{i-1}/x_i)} \ln (x/x_i) \,,  \quad {\rm for}  \quad x_i \leq x \leq x_{i-1}  \,,
\label{eq:fit1}
\end{equation}
where $y_i=S_{Pb}(x_i)$, $1 \leq i \leq 5$, are 5 free parameters of the fit. The fit function is continuous by construction.
Note that we did not attempt to use a more elaborate and smooth form of the fitting function to preserve the simplicity of its interpretation since we
concentrate on the general trend of the $x$-dependence (energy dependence) of $S_{Pb}(x)$.

To examine the small-$x$ asymptotic of $S_{Pb}(x)$, we have also considered a more constrained scenario, when the fit function is constant in the lowest-$x$ interval, 
$x \leq 10^{-4}$, but otherwise has the same form as in Fit 1,
\begin{equation}
S_{Pb}(x)_{|\rm{Fit\, 2}}=S_{Pb}(x;y_5=y_4,y_{i\leq 4})_{|\rm{Fit\, 1}} \,.
\label{eq:fit2}
\end{equation}
Fit 2 has 4 free parameters $y_i$ with $1 \leq i \leq 4$.

Further, to test sensitivity to the small-$x$ region, we combined the two lowest-$x$ intervals $10^{-5} < x < 10^{-4}$ and $10^{-4} <x < 10^{-3}$ into one, $10^{-5} < x < 10^{-3}$, and assumed the same piece-wise form as in the previous two cases.
 The corresponding fit functions are
\begin{equation}
S_{Pb}(x)_{|\rm{Fit\, 3}}=S_{Pb}(x;y_5=2y_4-y_3,y_{i\leq 4})_{|\rm{Fit\, 1}} \,,
\label{eq:fit3}
\end{equation}
and 
\begin{equation}
S_{Pb}(x)_{|\rm{Fit\, 4}} = S_{Pb}(x;y_5=y_4=y_3,y_{i\leq 3})_{|\rm{Fit\, 1}}  \,.
\label{eq:fit4}
\end{equation}
The resulting Fit 3 and Fit 4 have 4 and 3 free parameters, respectively. 

We determine the free fit parameters using the available UPC data on coherent 
$J/\psi$ photoproduction described above. In particular, using Eqs.~(\ref{eq:R}), (\ref{eq:cs_IA}), and (\ref{eq:upc_cs_2}), the GKSZ parametrization
of the $d\sigma^{\gamma+p \to J/\psi+p}(W_{\gamma p},t=0)/dt$ cross section on the proton described in Sec.~\ref{subsec:proton}, and the photon flux discussed
in Sec.~\ref{subsec:def}, we perform the $\chi^2$ fit to the UPC data~\cite{James:1975dr}.
We consider two options: in option $a$, we use only
the Run 1 and Run 2 data on $d\sigma^{AA \to J/\psi AA}/dy$ (21 points), and in option $b$ we combine all the data including the CMS and ALICE data on $\sigma^{\gamma A \to J/\psi A}(W_{\gamma p})$ (36 points). In the following, we corresponding fit results carry the labels $a$ and $b$, respectively.

Note that the STAR data point~\cite{STAR:2023gpk} enters out fit in the form of the nuclear suppression factor $S_{Pb}(x)$ at the 
average $\langle x \rangle =0.015$, which we determined using the STAR result for $S_{\rm coh}^{\rm Au}$,
\begin{equation}
S_{Pb}(\langle x \rangle =0.015)=\sqrt{S_{\rm coh}^{\rm Au} \frac{d\sigma^{\gamma+p \to J/\psi+p}(\langle W_{\gamma p}\rangle,t=0)/dt_{|\rm H1}}{d\sigma^{\gamma+p \to J/\psi+p}(\langle W_{\gamma p} \rangle,t=0)/dt_{|\rm GKSZ}}}=0.89 \pm 0.067 \,.
\label{eq:R_Au}
\end{equation}
In addition to taking the square root, we also corrected the STAR suppression factor by the ratio of the H1 and GKSZ parametrizations of the 
$d\sigma^{\gamma+p \to J/\psi+p}(W_{\gamma p},t=0)/dt$ differential cross sections at $\langle W_{\gamma p} \rangle=M_{J/\psi}/\sqrt{\langle x \rangle}$ since it has been defined with respect to the H1 parametrization~\cite{STAR:2023gpk}, while our $\chi^2$ analysis as well as the CMS~\cite{CMS:2023snh} and ALICE~\cite{ALICE:2023jgu}
analyses use the GKSZ fit as the baseline.
We also neglected a small difference between the nuclear suppression factors for $^{208}$Pb and $^{197}$Au since it is much smaller than the experimental 
uncertainty in Eq.~(\ref{eq:R_Au}) because the nuclear suppression for such close nuclei is expected to be very 
similar.

Table~\ref{table:chi2} summarized the resulting values of $\chi^2$. One can see from the Table that while all 4 parametrizations can fit
the $d\sigma^{AA \to J/\psi AA}/dy$ data, Fit 4 with the flat $x$-dependence for $x < 10^{-3}$ fails to fit $\sigma^{\gamma A \to J/\psi A}(W_{\gamma p})$. Hence, one can conclude that these data constrain the small-$x$ behavior of $S_{Pb}(x)$.
Note that the use of the H1 parametrization for $d\sigma^{\gamma+p \to J/\psi+p}(W_{\gamma p},t=0)/dt$ results in very similar values 
of $\chi^2$.

\begin{table}[h]
\begin{center}
\begin{tabular}{|c|c|c|}
\hline
& Option $a$ & Option $b$ \\
 & (21 points) & (36 points) \\
\hline
Fit 1 & 12.8 & 34.9\\
Fit 2 & 12.9 & 36.9 \\
Fit 3 & 12.8 & 35.0 \\
Fit 4 & 15.4 & 95.1\\
\hline
\end{tabular}
\end{center}
\caption{The $\chi^2$ values for the four fits of the nuclear suppression factor $S_{Pb}(x)$ using the data on $d\sigma^{AA \to J/\psi AA}/dy$ (21 points) and 
$d\sigma^{AA \to J/\psi AA}/dy$ and $\sigma^{\gamma A \to J/\psi A}(W_{\gamma p})$ (36 points) for coherent $J/\psi$
photoproduction in heavy-ion UPCs.
}
\label{table:chi2}
\end{table}

\begin{figure}[t]
\centerline{%
\includegraphics[width=9cm]{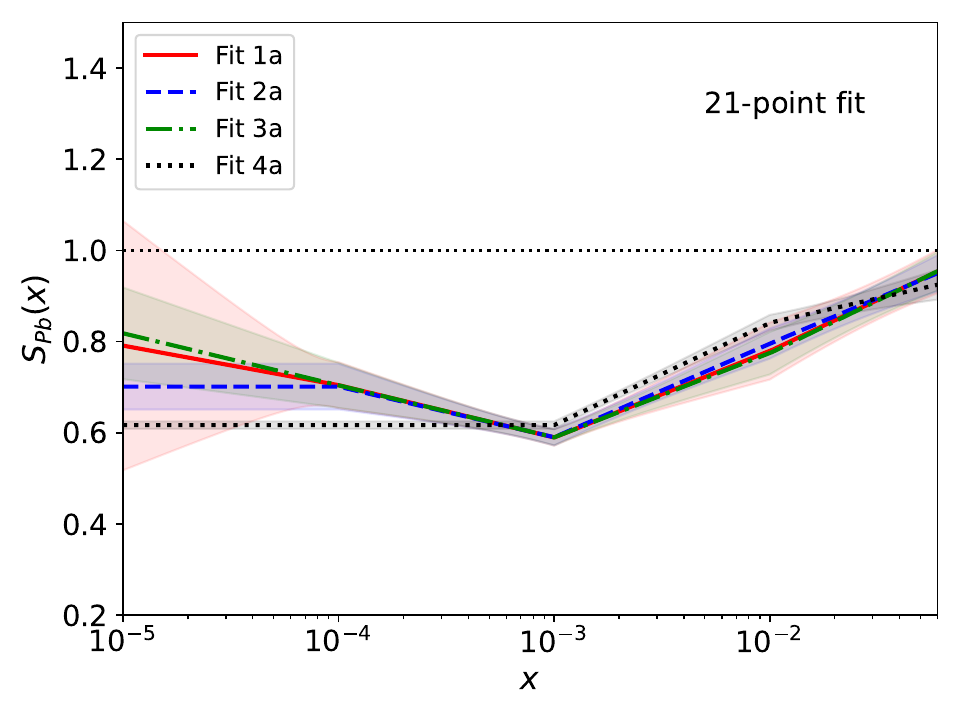}
\includegraphics[width=9cm]{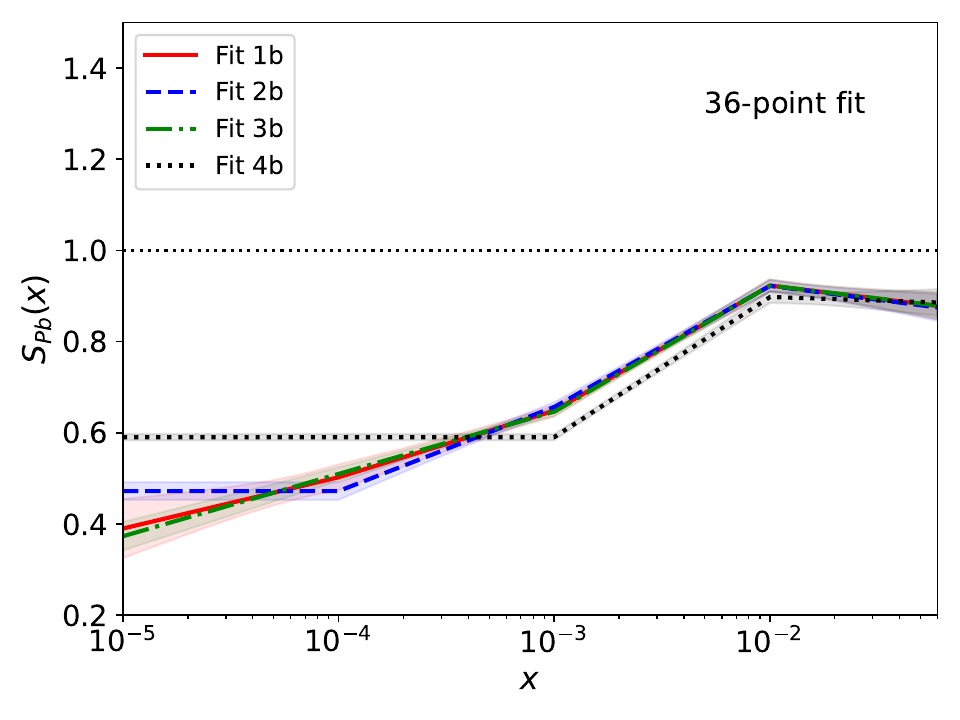}
}
\caption{Four fit functions with uncertainties for the nuclear suppression factor $S_{Pb}(x)$ as a function of $x$, which are obtained by performing the $\chi^2$ fit to the data on $d\sigma^{AA \to J/\psi AA}/dy$ (21 points, left panel) and 
$d\sigma^{AA \to J/\psi AA}/dy$ and $\sigma^{\gamma A \to J/\psi A}(W_{\gamma p})$ (36 points, right panel) for coherent $J/\psi$
photoproduction in heavy-ion UPCs.
}
\label{fig:fit}
\end{figure}

Figure~\ref{fig:fit} shows Fits 1-4 as a function of $x$, where the shaded bands give their uncertainties evaluated using
uncorrelated uncertainties of the fit parameters; the left and right panels corresponds to the fits performed using 21 (option $a$) and 36 
(option $b$) data points,
respectively. 
The fit solutions presented in these two panels are quite distinct. 
One can see in the left panel that while the $d\sigma^{AA \to J/\psi AA}/dy$ data allow us to determine $S_{Pb}(x)$ for $x \geq 10^{-3}$ rather well,
they do not constrain the $x < 10^{-3}$ region. In particular, the data do not distinguish between the flat and rising behavior
of the fitting functions at small $x$. Also, the uncertainty band for Fit 1a becomes very large for $x < 10^{-4}$.  

At the same time, one can see in the right panel of Fig.~\ref{fig:fit} that the combination of the $d\sigma^{AA \to J/\psi AA}/dy$
and $\sigma^{\gamma A \to J/\psi A}(W_{\gamma p})$ data provides rather tight constraints in a wide range of $x$, $10^{-4} \leq x \leq 0.05$.
In particular, the successful fits (Fits 1-3) favor $S_{Pb}(x)$, which decreases with a decrease of $x$ in the $10^{-4} \leq x \leq 10^{-2}$ interval.
The behavior for smaller values of $x$, $x < 10^{-4}$, is less constrained and is consistent with both the continuing decrease 
of $S_{Pb}(x)$ (Fits 1b and 3b) 
or with a constant value of $S_{Pb}(x)$ for $x < 10^{-4}$ (Fit 2b).

Note that we have also performed a separate fit to the 15 data points on $\sigma^{\gamma A \to J/\psi A}(W_{\gamma p})$ omitting 
the $d\sigma^{AA \to J/\psi AA}/dy$ data. For $x < 0.01$, the resulting fit functions turned out to be similar to those shown in the right panel of Fig.~\ref{fig:fit}, but with wider uncertainty bands. For $x \geq 0.01$, the fit functions first exceed their Fig.~\ref{fig:fit} counterparts at $x \approx 0.01$ and then dip below them for $x > 0.03$, which leads to a certain tension with the $d\sigma^{AA \to J/\psi AA}/dy$ data, especially with the LHCb data at $|y| \approx 2-3$.

\subsection{Comparison with data on coherent $J/\psi$ photoproduction in heavy-ion UPCs}
\label{subsec:fit_comp}

It is instructive to examine how different fit solutions for $S_{Pb}(x)$ affect the magnitude and shape of the $d\sigma^{AA \to J/\psi AA}/dy$
and $\sigma^{\gamma A \to J/\psi A}(W_{\gamma p})$ cross sections and their comparison to the data used in the fits. 

\begin{figure}[t]
\centerline{%
\includegraphics[width=9cm]{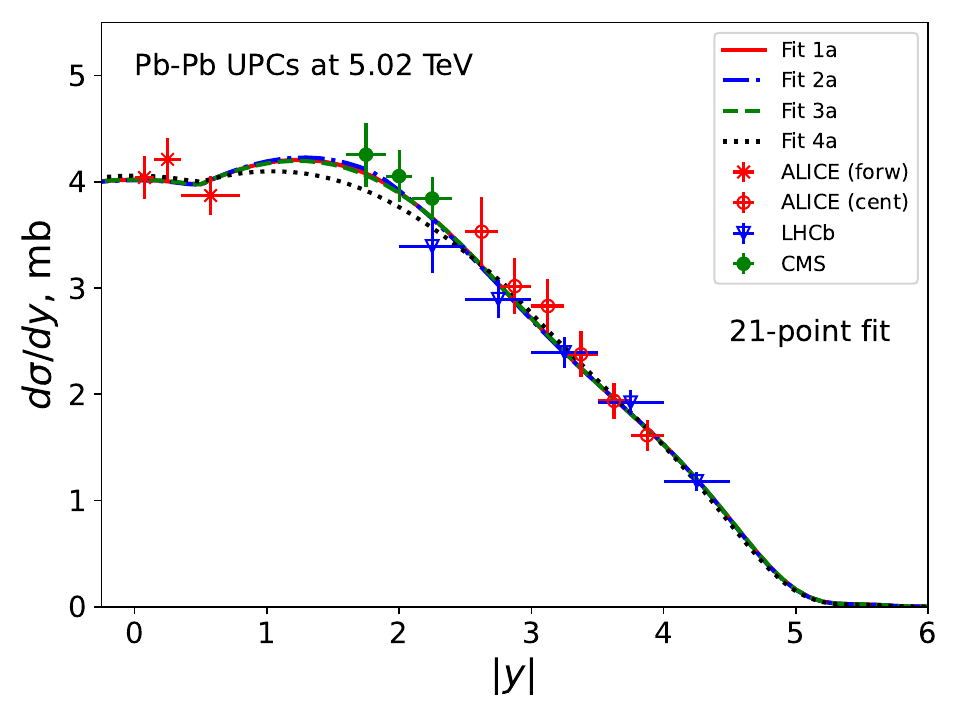}
\includegraphics[width=9cm]{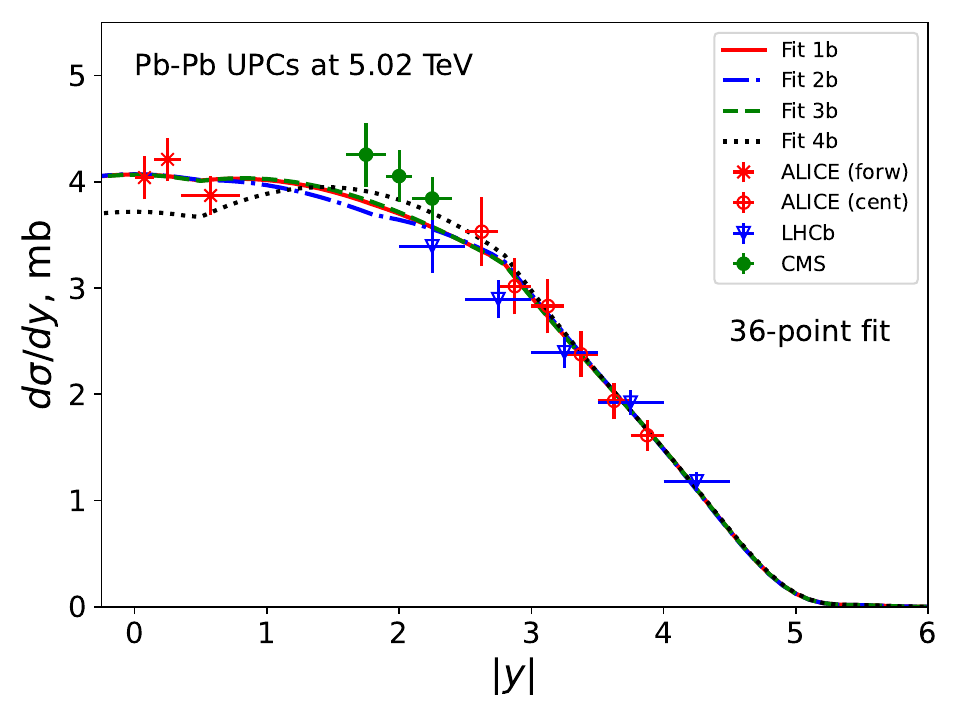}
}
\caption{The rapidity-differential  $d\sigma^{AA \to J/\psi AA}/dy$ cross section of coherent $J/\psi$ photoproduction
in Pb-Pb UPCs at 5.02 TeV as a function of the $J/\psi$ rapidity $|y|$. Predictions of Eq.~(\ref{eq:upc_cs_2}) with $S_{Pb}(x)$ 
given by the four fits (option $a$ in the left panel and option $b$ in the right panel) are compared with the ALICE~\cite{ALICE:2021gpt,ALICE:2019tqa}, LHCb~\cite{LHCb:2022ahs}, and CMS~\cite{CMS:2023snh} data.
}
\label{fig:ydep_fit}
\end{figure}

Figure~\ref{fig:ydep_fit} presents the rapidity-differential $d\sigma^{AA \to J/\psi AA}/dy$ cross section of coherent $J/\psi$ photoproduction
in Pb-Pb UPCs at 5.02 TeV as a function of the $J/\psi$ rapidity $|y|$ (for symmetric $AA$ UPCs, the rapidity distribution is symmetric). 
The theory curves labeled ``Fit 1, 2, 3, 4'' are obtained using Eq.~(\ref{eq:upc_cs_2}), where $S_{Pb}(x)$ is given by the fits described above;
the left and right panels correspond to the fits using 21 (option $a$) and 36 (option $b$) data points, respectively. 
They are compared with the experimental data that have been used in the fits, which include those by
ALICE at central rapidities~\cite{ALICE:2021gpt} (red crosses), ALICE at forward rapidities~\cite{ALICE:2019tqa} (red open circles), LHCb at forward rapidities~\cite{LHCb:2022ahs} (blue inverted triangles), and CMS at intermediate rapidity~\cite{CMS:2023snh} (green solid circles).

One can see from this figure that Fits 1, 2, and 3 correspond to the essentially indistinguishable description of the data in the entire range of $y$. Fit 4b in the case of 36 fitted data points lies approximately 10\% 
below the other predictions at $y \approx 0$, but agrees with them for $ |y| > 1$. Note, however, that this 
fit is excluded because of the very large $\chi^2$, see Table~\ref{table:chi2}. This comparison supports our conclusion that the $d\sigma^{AA \to J/\psi AA}/dy$ data alone do not constrain the nuclear modification factor 
$S_{Pb}(x)$ for $x < 10^{-3}$ well enough.

\begin{figure}[t]
\centerline{%
\includegraphics[width=9cm]{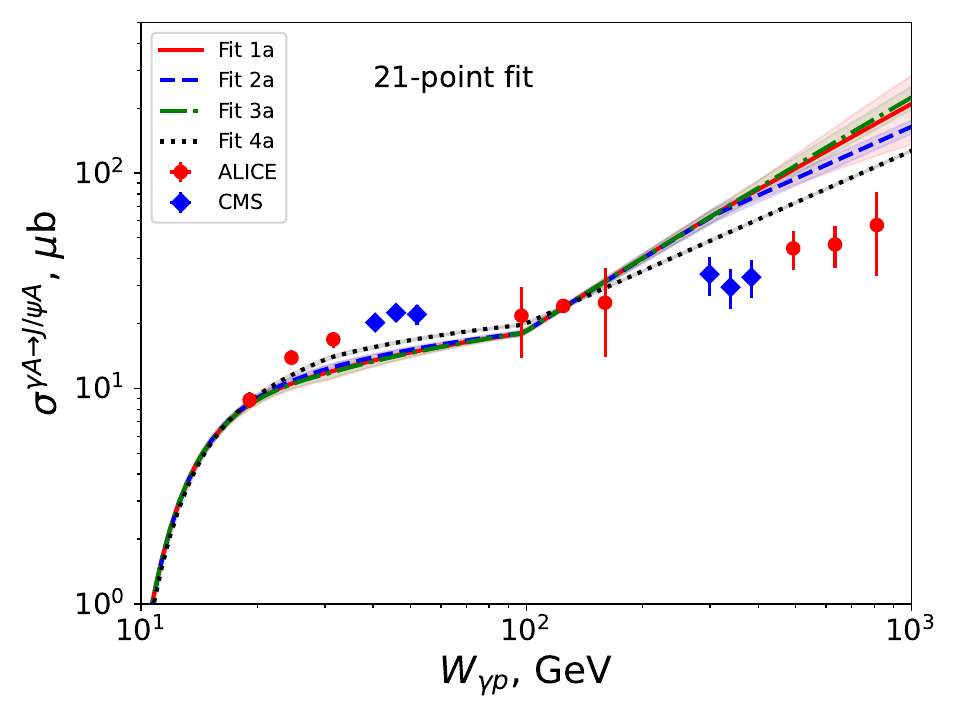}
\includegraphics[width=9cm]{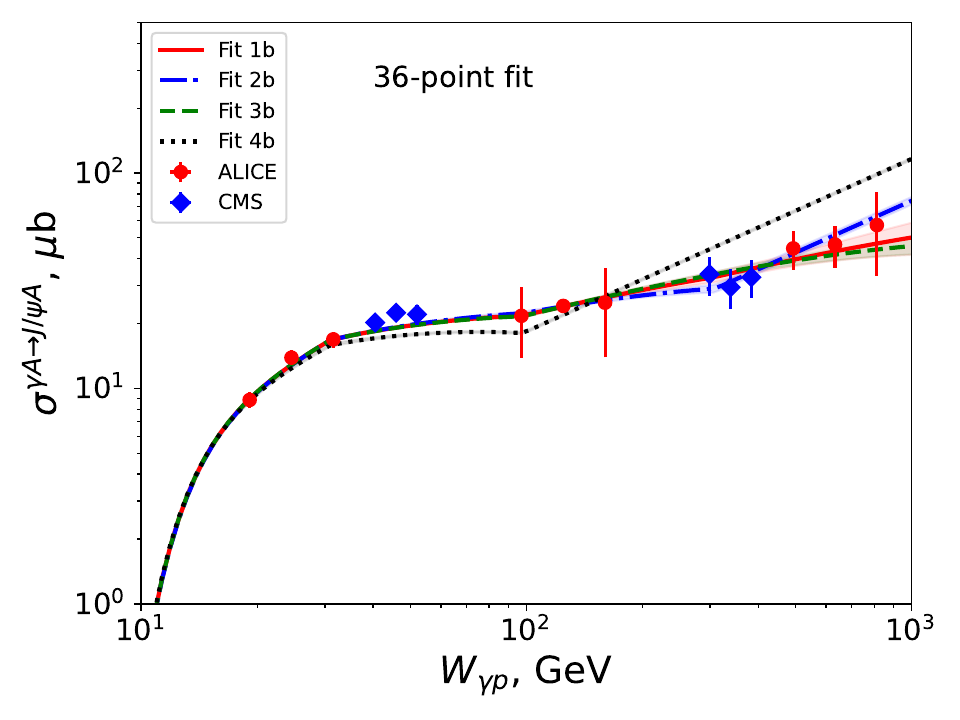}
}
\caption{The nuclear photoproduction cross section $\sigma^{\gamma A \to J/\psi A}(W_{\gamma p})$
as a function of $W_{\gamma p}$. Predictions of Eq.~(\ref{eq:R}), where $S_{Pb}(x)$  is
given by the four fits (option $a$, left panel and option $b$, right panel), are compared with the CMS~\cite{CMS:2023snh} and ALICE~\cite{ALICE:2023jgu} data.
}
\label{fig:Wdep_fit}
\end{figure}

In Fig.~\ref{fig:Wdep_fit}, we show the nuclear photoproduction cross section $\sigma^{\gamma A \to J/\psi A}(W_{\gamma p})$
as a function of $W_{\gamma p}$, where the results of Eq.~(\ref{eq:R}) using the four fits for $S_{Pb}(x)$ are compared with the CMS~\cite{CMS:2023snh} (blue solid diamonds) and ALICE~\cite{ALICE:2023jgu} (red solid circles) data. The shaded bands, which are visible only for $W_{\gamma p} > 300$ GeV, quantify the propagated 
uncertainty of the fit parameters.
One can see from the left panel of this figure that 
the 21-point fits (option $a$), which do not include the data on $\sigma^{\gamma A \to J/\psi A}(W_{\gamma p})$, do not reproduce them,
except for three ALICE points around $W_{\gamma p} = 120$ GeV, which are consistent with the ALICE data on $d\sigma^{AA \to J/\psi AA}/dy$
at central rapidities~\cite{ALICE:2021gpt} 
(it is effectively the same data, but presented in a different way),
and the smallest-$W_{\gamma p}$ ALICE data point.

At the same time, the right panel of Fig.~\ref{fig:Wdep_fit} demonstrates that Fits 1b, 2b, and 3b can successfully accommodate
the data on both $d\sigma^{AA \to J/\psi AA}/dy$ and $\sigma^{\gamma A \to J/\psi A}(W_{\gamma p})$.
It is important to note that within sizable experimental errors for $W_{\gamma p} > 100$ GeV, the data cannot distinguish 
between Fits 1b and 3b with a decreasing $S_{Pb}(x)$ for small $x$ and Fit 2b with a flat $S_{Pb}(x)$ for $x < 10^{-4}$.
As we discussed above, Fit 4b with a flat $x$-behavior of $S_{Pb}(x)$ for $x < 10^{-3}$ is inconsistent with the data 
for $W_{\gamma p} > 120$ GeV.

\begin{figure}[t]
\centerline{%
\includegraphics[width=9cm]{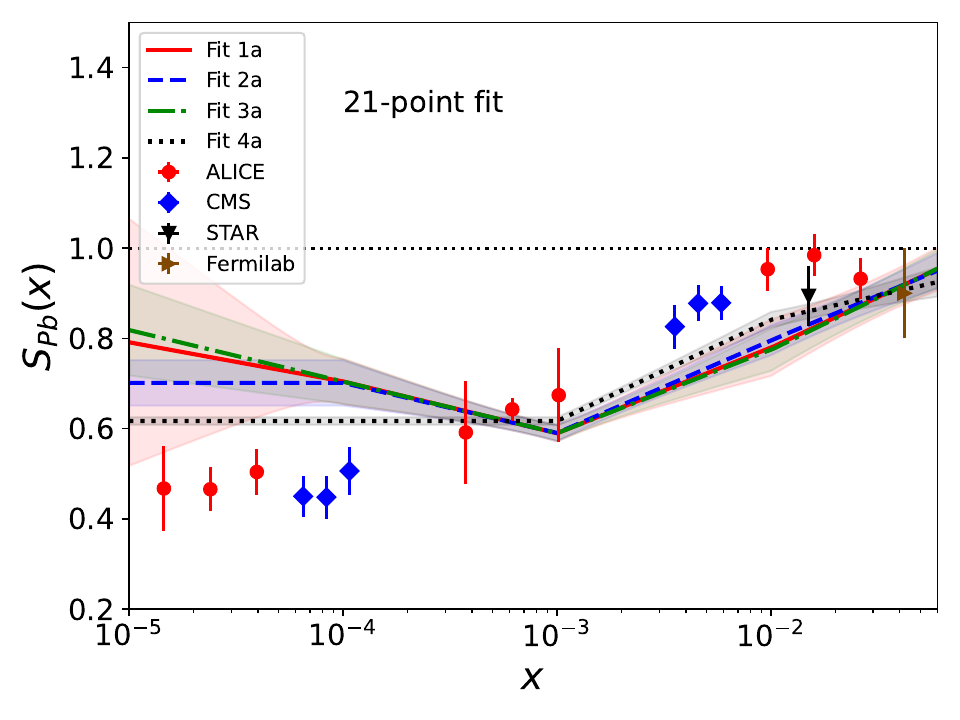}
\includegraphics[width=9cm]{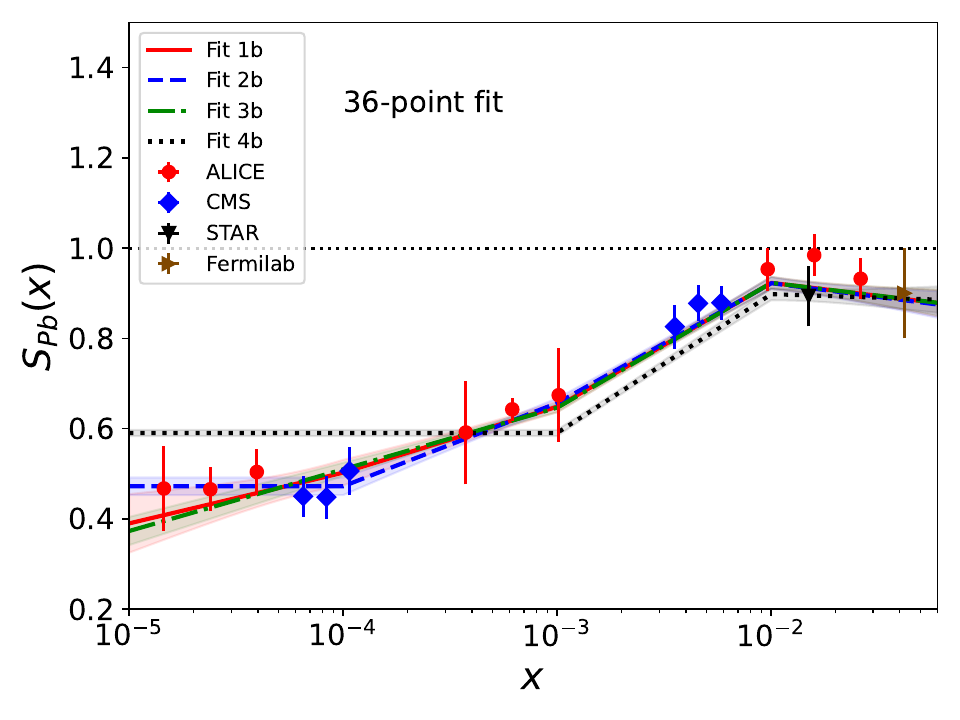}
}
\caption{The nuclear suppression factor $S_{Pb}(x)$ as a function of $x$. The results of the four fits with uncertainty bands
are compared with the CMS~\cite{CMS:2023snh}, ALICE~\cite{ALICE:2023jgu}, and STAR~\cite{STAR:2023gpk} data as well as with
the value extracted from Fermilab data~\cite{Guzey:2020ntc}. The left and right panels correspond respectively to the 21-point (option $a$) 
and 36-point (option $b$) fits.
}
\label{fig:Rg_fit}
\end{figure}

Finally, the nuclear suppression factor $S_{Pb}(x)$ determined as a result of our $\chi^2$ analysis
can be directly compared with the nuclear suppression factor extracted by CMS~\cite{CMS:2023snh} and ALICE~\cite{ALICE:2023jgu} collaborations. 
The procedure used in these analyses follows that one proposed in Ref.~\cite{Guzey:2013xba} and is fully consistent with our definitions in 
Sec.~\ref{subsec:def}.
Figure~\ref{fig:Rg_fit} presents a comparison of the four fits for $S_{Pb}(x)$ as a function of $x$ with the CMS~\cite{CMS:2023snh} 
(blue solid diamonds), ALICE~\cite{ALICE:2023jgu} (red solid circles), and STAR (black inverted triangle), see Eq.~(\ref{eq:R_Au}),
 values. In addition, we also show by the brown right triangle the value of $S_{Pb}(x)$ determined in~\cite{Guzey:2020ntc} from the Fermilab data on
 inclusive $J/\psi$ production by a 120-GeV photon beam on fixed nuclear targets~\cite{FermilabTaggedPhotonSpectrometer:1986xzf}, 
 \begin{equation}
 S_{Pb}(\langle x \rangle=0.042)=0.90 \pm 0.10 \,.
 \label{eq:R_Fermilab}
 \end{equation}
 Note that the curves in Fig.~\ref{fig:Rg_fit} are the same as in Fig.~\ref{fig:fit} with 
 the shaded bands giving the uncertainty of $S_{Pb}(x)$ due to uncertainties 
of the fit parameters; the left and right panels correspond to the 21-point and 36-point fits, respectively.

The conclusions that one can draw from the comparison presented in Fig.~\ref{fig:Rg_fit} is the same as that for Fig.~\ref{fig:Wdep_fit}:
Option $a$ fits fail to reproduce the data at very small $x$ and large $x$, with the exception of the three ALICE points at $x \approx 10^{-3}$,
the ALICE point at $x=0.026$, and the STAR point at $x=0.015$ (used in the fits). Note that the Fermilab value with a significant experimental
uncertainty fits the pattern and is also reproduced by all the fits.

In contrast, as one can see in the right panel of Fig.~\ref{fig:Rg_fit}, Fits 1b, 2b, and 3b describe very well all the data points shown. 
Additionally, within large experimental and theoretical fit uncertainties for $x < 10^{-4}$, both Fit 1b and Fit 3b with a decreasing $S_{Pb}(x)$
and Fit 2b with a constant  $S_{Pb}(x)$ for $x < 10^{-4}$ can be accommodated equally well by the data.
Fit 4b with a constant $S_{Pb}(x)$ for $x < 10^{-3}$ is ruled out by the data.

Note that the results presented in this section are obtained using the GKSZ parametrization for $d\sigma^{\gamma+p \to J/\psi+p}(W,t=0)/dt$, see Sec.~\ref{subsec:proton}.
We have checked that the use of the $d\sigma^{\gamma+p \to J/\psi+p}(W,t=0)/dt_{|\rm H1}$ parametrization results in a very similar quality of the fit in all the considered cases. The only difference is the normalization and to a much lesser degree the shape of the fitting function, which becomes rescaled by the factor
\begin{equation}
\left[\frac{d\sigma^{\gamma+p \to J/\psi+p}(W_{\gamma p},t=0)/dt_{|\rm GKSZ}}{d\sigma^{\gamma+p \to J/\psi+p}(W_{\gamma p},t=0)/dt_{|\rm H1 }}\right]^{1/2} \leq 1 \,.
\label{eq:scale}
\end{equation}
This rescaling does not affect the results for $d\sigma^{AA \to J/\psi AA}/dy$ in Fig.~\ref{fig:ydep_fit} and
$\sigma^{\gamma A \to J/\psi A}(W_{\gamma p})$ in Fig.~\ref{fig:Wdep_fit}. It only decreases the values of $S_{Pb}(x)$ 
for $x > 10^{-3}$ by $5-10$\% (the effect increases with an increase of $x$) without modifying $S_{Pb}(x)$ for $x < 10^{-3}$ in Figs.~\ref{fig:fit}
and \ref{fig:Rg_fit}.

Note, however, that despite the freedom in the choice of the parametrization for $d\sigma^{\gamma+p \to J/\psi+p}(W,t=0)/dt$, which is compensated
by appropriate modifications of $S_{Pb}(x)$ in the calculations of $d\sigma^{AA \to J/\psi AA}/dy$ and
$\sigma^{\gamma A \to J/\psi A}(W_{\gamma p})$, 
it is important to use the same baseline fit for $d\sigma^{\gamma+p \to J/\psi+p}(W,t=0)/dt$
in the theoretical and experimental analyses of the nuclear suppression factor $S_{Pb}(x)$.  As we explained above, the use of inconsistent parametrizations
induces a $5-10$\% correction that has nothing to do with nuclear modifications.

\section{Nuclear suppression from leading twist nuclear shadowing}
\label{sec:lta}

\subsection{Leading twist approach to small-$x$ nuclear PDFs}
\label{subsec:lta_theory}

The leading twist approach (LTA) to nuclear shadowing~\cite{Frankfurt:2011cs} makes definite predictions for the ratio of the gluon distributions in a nucleus $g_A(x,Q^2)$ and the free proton $g_p(x,Q^2)$, $R_g(x,Q^2)=g_A(x,Q^2)/[Ag_p(x,Q^2)]$,
which can be readily converted into the nuclear suppression factor
for coherent $J/\psi$ photoproduction in heavy-ion UPCs~\cite{Guzey:2013xba,Guzey:2013qza}. It combines features of the Gribov-Glauber model for hadron-nucleus scattering at high energies with QCD factorization theorems for inclusive deep inelastic scattering (DIS)~\cite{CTEQ:1993hwr} and hard diffraction~\cite{Collins:1997sr}. As a result,  it allows one to calculate $R_g(x,Q^2)$ in terms of the gluon diffractive PDF of the proton $g_p^{D(4)}$ 
as a function of $x$ at some input scale $Q_0^2={\cal O}(\rm few)$ GeV$^2$ chosen to minimize possible higher-twist corrections in diffractive DIS; the subsequent $Q^2$ dependence of $R_g(x,Q^2)$ is given by leading-twist Dokshitzer-Gribov-Lipatov-Altarelli-Parisi (DGLAP) evolution equations.

The LTA prediction for $R_g(x,Q_0^2)$ has the following form
\begin{eqnarray}
R_g(x,Q_0^2)&=&\frac{g_A(x,Q_0^2)}{Ag_p(x,Q_0^2)} = 1-8 \pi \Re e \frac{(1-i \eta)^2}{1+\eta^2} \int_{x}^{x_0} dx_{\Pomeron} \frac{\beta g_p^{D(4)}(x,x_{\Pomeron},t=0,Q_0^2)}{A xg_p(x,Q_0^2)} \nonumber\\
&\times& \int d^2 \vec{b} \int^{\infty}_{-\infty} dz_1 \int_{z_1}^{\infty} dz_2 \rho_A(\vec{b},z_1) \rho_A(\vec{b},z_2)e^{i (z_1-z_2) x_{\Pomeron}m_N} e^{-\frac{1-i \eta}{2} \sigma_{\rm soft}(x,Q_0^2) \int_{z_1}^{z_2} dz^{\prime} \rho_A(\vec{b},z^{\prime})} \,,
\label{eq:g_A_lta}
\end{eqnarray}
where $\rho_A(\vec{r})$ is the nuclear density~\cite{DeVries:1987atn} normalized to the number of nucleons, $\int d^3 \vec{r} \rho_A(\vec{r})=A$;
$z_{1,2}$ and $z^{\prime}$ and $\vec{b}$ are the longitudinal and transverse coordinates of the target nucleons, respectively. Note that since the slope of 
the $t$ dependence of the proton diffractive structure function measured at HERA, $B_{\rm diff} \approx 6$ GeV$^{-2}$~\cite{H1:2006uea}, is much smaller than 
the nucleus radius squared, $g_p^{D(4)}$ is evaluated at $t=0$ and all nucleons are located at the same $\vec{b}$. The orderings of integration over $z$ take into account the space-time development of virtual photon-nucleus scattering in the target rest frame; the phase factor $e^{i (z_1-z_2) x_{\Pomeron}m_N}$
with $m_N$ the nucleon mass originates from the non-zero longitudinal momentum transfer in the $\gamma^{\ast}+N \to X +N$ diffractive scattering
on target nucleons ($X$ is the diffractively-produced final state).

The diffractive PDF $g_p^{D(4)}(x,x_{\Pomeron},t=0,Q_0^2)$ depends on two longitudinal momentum fractions $x$ and $x_{\Pomeron}$, where the latter represents
the momentum fraction carried by the diffractive exchange (commonly called the Pomeron), the momentum transfer $t$, and the factorization scale $Q_0$. 
To simply their analysis, diffractive PDFs are usually presented as a product of the Pomeron flux depending on $x_{\Pomeron}$ and $t$ and 
parton distributions inside the Pomeron, which are functions of the momentum fraction $\beta=x/x_{\Pomeron}$ and $Q_0$.
In our calculations, we employ diffractive PDFs obtained by the H1 collaboration using a QCD analysis of data on inclusive diffraction in DIS at HERA~\cite{H1:2006zyl}.
Note that other global QCD analyses of diffractive PDFs available in the literature~\cite{ZEUS:2009uxs,Goharipour:2018yov,Salajegheh:2023jgi}
agree with the H1 fit.

In Eq.~(\ref{eq:g_A_lta}), integration over $x_{\Pomeron}$ runs from the lowest value allowed by kinematics and up to $x_0=0.1$ determined by the usual condition on the produced diffractive
masses $M_X$, $M_X^2/W_{\gamma p}^2 \leq 0.1$. However, the exact numerical value of $x_0$ plays a minor role because the integrand in Eq.~(\ref{eq:g_A_lta})
for $x_{\Pomeron}$ close to $x_0$ is suppressed by the rapidly oscillating $e^{i (z_1-z_2) x_{\Pomeron}m_N}$ term.   

\begin{figure}[t]
\centerline{%
\includegraphics[width=10cm]{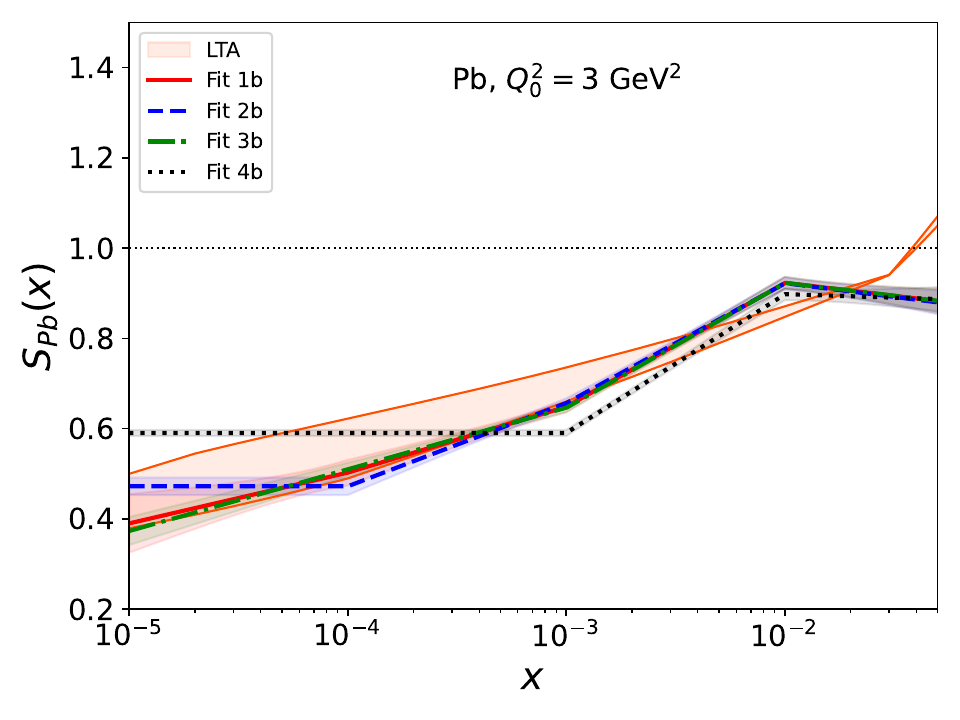}}
\caption{The nuclear suppression factor $S_{Pb}(x)$ as a function of $x$. The LTA predictions for $R_g(x,Q_0^2)$ for Pb at $Q_0^2=3$ GeV$^2$ given by the shaded band are compared with the
Fit 1b-4b results.
}
\label{fig:Comp_Rg_2024_all}
\end{figure}

The second exponential factor in Eq.~(\ref{eq:g_A_lta}) is a result of eikonalization of the interaction with $N \geq 3$ nucleons of the target, whose 
effective cross section is $\sigma_{\rm soft}(x,Q_0^2)$. It depends on the parton flavor, but assumed to be independent of $M_X$ or $x_{\Pomeron}$.
Its numerical value (lower and upper limits) can be estimated using plausible models for the hadronic component of the virtual photon, see details in Ref.~\cite{Frankfurt:2011cs}. 
For instance, for $Q_0^2=3$ GeV$^2$, $\sigma_{\rm soft}(x,Q_0^2)=26-45$ mb at $x=10^{-3}$ and $\sigma_{\rm soft}(x,Q_0^2)=30-52$ mb at $x=10^{-4}$.
The theoretical uncertainty of $\sigma_{\rm soft}(x,Q_0^2)$ propagates into the uncertainty of LTA predictions for the 
$R_g(x,Q_0^2)$ ratio shown by the shaded band in Fig.~\ref{fig:Comp_Rg_2024_all}.

Finally, $\eta$ is the ratio of the real to imaginary parts of the $\gamma^{\ast}+N \to X +N$ diffractive amplitude on target nucleons. It 
can be estimated using the $x_{\Pomeron}$ dependence of $g_p^{D(4)}$, $\eta \approx \pi/2 (\alpha_{\Pomeron}(0)-1)=0.17$, where  
$\alpha_{\Pomeron}(0)=1.111$~\cite{H1:2006zyl} is the intercept of the effective Pomeron trajectory. 

In the limit of small $x$, $x < 10^{-2}$, Eq.~(\ref{eq:g_A_lta}) can be simplified by omitting the $e^{i (z_1-z_2) x_{\Pomeron}m_N}$ term. 
After integration over $z_1$ and $z_2$ using that 
 $\int_{z_1}^{\infty} dz_2 \rho_A(\vec{b},z_2) e^{-L\int_{z_1}^{z_2} dz^{\prime} \rho_A(\vec{b},z^{\prime})}=
 (1-e^{-L \int^{\infty}_{z_1} dz^{\prime} \rho_A(\vec{b},z^{\prime})})/L$ and $\int_{-\infty}^{\infty} dz_1 \rho_A(\vec{b},z_1) e^{-L\int_{z_1}^{\infty} dz^{\prime} \rho_A(\vec{b},z^{\prime})}=
 (1-e^{-L T_A(\vec{b})})/L$, where $L=\frac{1-i \eta}{2} \sigma_{\rm soft}(x,Q_0^2)$,
 one obtains
\begin{equation}
R_g(x,Q_0^2)=1-\frac{\sigma_2(x,Q_0^2)}{\sigma_{\rm soft}(x,Q_0^2)}+2 \frac{\sigma_2(x,Q_0^2)}{A\sigma_{\rm soft}^2(x,Q_0^2)} \Re e \int d^2 \vec{b}
\left(1-e^{-\frac{1-i \eta}{2} \sigma_{\rm soft}(x,Q_0^2) T_A(\vec{b})} \right) \,,
\label{eq:g_A_lta_2}
\end{equation}
where $T_A(\vec{b})=\int^{\infty}_{-\infty} dz \rho_A(\vec{b},z)$ is the nuclear optical density. In Eq.~(\ref{eq:g_A_lta_2}), we introduced the cross section $\sigma_2(x,Q_0^2)$,
which quantifies the relative probability of hard diffraction in the gluon channel,
\begin{equation}
\sigma_2(x,Q_0^2)=\frac{16 \pi}{(1+\eta^2)} \int_{x}^{x_0} dx_{\Pomeron} \frac{\beta g_p^{D(4)}(x,x_{\Pomeron},t=0,Q_0^2)}{xg_p(x,Q_0^2)} \,.
\label{eq:sigma_2}
\end{equation}
In this notation, Eq.~(\ref{eq:g_A_lta_2}) has a transparent physical interpretation. The suppression of $R_g(x,Q_0^2) < 1$ is caused by nuclear shadowing, which 
arises as an effect of multiple scattering of hadronic components of the virtual photon off target nucleons. These components are modeled as superposition of 
a state interacting with a vanishingly small cross section and whose probability is $\lambda=1-\sigma_2(x,Q_0^2)/\sigma_{\rm soft}(x,Q_0^2) \leq 1 $, and a state
interacting with the cross section $\sigma_{\rm soft}(x,Q_0^2)$ and whose probability is $1-\lambda$.
The nuclear cross section for the latter component has the standard form of the Glauber model.
As can be seen by expanding Eq.~(\ref{eq:g_A_lta_2}) in powers of $T_A(\vec{b})$, the interaction with $N=2$ nucleons is given by $\sigma_2(x,Q_0^2)$, while
the interaction with $N \geq 3$ nucleons is determined by $\sigma_{\rm soft}(x,Q_0^2)$.

While LTA offers a viable alternative to global QCD analyses of nuclear PDFs, it does not naturally contain the nuclear enhancement (antishadowing) of the gluon distribution around $x_0$ present in modern nuclear PDFs~\cite{Eskola:2021nhw,Kovarik:2015cma,AbdulKhalek:2022fyi}.
Hence, the gluon antishadowing demands a separate modeling, which is realized in LTA by requiring conservation of the momentum sum rule for nuclear PDFs~\cite{Frankfurt:2011cs}. It relies on the result of~\cite{Frankfurt:1990xz}, where the evidence for an enhancement of the nuclear
gluon distribution was found using an analysis of high-precision data on DIS on nuclear targets and the QCD baryon number and momentum sum rules.
In particular, assuming that the gluon antishadowing affects $R_g(x,Q_0^2)$ only in the interval $0.03 \leq x \leq 0.2$ at the initial scale $Q_0$, where it is parametrized in the following
simple form, 
\begin{equation}
R_g^{\rm anti}(x,Q_0^2)= N_{\rm anti} (x-0.03) (0.2-x) \,,
\label{eq:anti}
\end{equation}
one can readily find the free parameter $N_{\rm anti}$ from the momentum sum rule. Its value somewhat depends on the choice of the baseline proton PDFs.
In the case of the leading-order CTEQ6L proton PDFs~\cite{Pumplin:2002vw} used in our analysis, we find that $N_{\rm anti}=26-30$. 
This range corresponds to the ``low shadowing'' and ``high shadowing'' LTA predictions, which are obtained using the upper and lower boundaries on
  $\sigma_{\rm soft}(x,Q_0^2)$. It leads to an approximately 20\% gluon antishadowing at $x=0.1$ at $Q_0^2=3$ GeV$^2$.
  
Note that the shape and $x$-support of the gluon antishadowing in Eq.~(\ref{eq:anti}) agrees with the result of the QCD analysis of scaling 
violations of high statistics NMC data on the ratio of the $F_{2}^{Sn}/F_{2}^C$ nuclear structure functions~\cite{Gousset:1996xt}.

The LTA predictions for $R_g(x,Q_0^2)=g_A(x,Q_0^2)/[Ag_p(x,Q_0^2)]$ for $^{208}$Pb as a function of $x$ are shown in
Fig.~\ref{fig:Comp_Rg_2024_all} by the shaded band. Its upper and lower boundaries correspond to the ``low shadowing'' and ``high shadowing'' predictions,
respectively, see the discussion above. They are compared with Fits 1b-4b for the nuclear suppression factor $S_{Pb}(x)$ discussed in Sec.~\ref{subsec:fit}.
One can see from the figure that apart from Fit 4b ruled out by the $\sigma^{\gamma A \to J/\psi A}(W_{\gamma p})$ data, the LTA and fit results agree within the
LTA uncertainties in the wide interval of $10^{-5} < x < 0.01$. For $x > 0.02$, the LTA and fit predictions begin to deviate from each other since the UPC data favor a continuing suppression $S_{Pb}(x) < 1$, while the momentum sum rule for nuclear PDFs requires an eventual enhancement of $R_g(x,Q_0^2) > 1$ due to
the assumed gluon antishadowing.

\subsection{Comparison of LTA predictions with data on coherent $J/\psi$ photoproduction in heavy-ion UPCs}
\label{subsec:lta_comp}

In the leading double logarithmic approximation of perturbative QCD and the static (non-relativistic) limit for the charmonium wave function,  
the differential cross section of exclusive $J/\psi$ photoproduction is proportional to the gluon density of the target squared~\cite{Ryskin:1992ui}.
Applying it to nuclear targets, one can relate the cross section of coherent $J/\psi$ photoproduction in heavy-ion UPCs to the factor $R_g(x,Q_0^2)$ quantifying nuclear modifications, primarily nuclear shadowing, of the gluon distribution in nuclei~\cite{Guzey:2013xba,Guzey:2013qza}. 
Note that in the framework of collinear factorization, this process is subject to very large next-to-leading order (NLO) perturbative corrections~\cite{Ivanov:2004vd}, which complicate interpretation of these UPC data in terms of gluon shadowing because of large cancellations 
in the sum of the leading-order (LO) and NLO gluon contributions and the emerging dominance of the quark contribution~\cite{Eskola:2022vpi,Eskola:2022vaf}.
It indicates instability of the perturbation series for this process at high energies, which is, however, could be restored using methods of small-$x$ 
resummation~\cite{Jones:2015nna,Jones:2016ldq}.

In the following, we ignore this issue and
use the LO formalism of~\cite{Guzey:2013xba,Guzey:2013qza}. 
It is generally expected that after the small-$x$ resummation, the NLO results should be close to the LO ones, which thus provide a useful benchmark for the considered process. This is further supported by the observation that the magnitude of nuclear effects
parameterized by $S_{Pb}(x)$ should be larger than the difference between the LO and resummed NLO results. 
Thus, our formalism 
relies on the gluon dominance of exclusive $J/\psi$ photoproduction
and allows one to identify the gluon shadowing factor $R_g(x,Q_0^2)$ discussed in Sec.~\ref{subsec:lta_theory} with the nuclear suppression factor
$S_{Pb}(x)$ introduced in Sec.~\ref{sec:suppression}. In particular, the $t$-integrated cross section of coherent $J/\psi$ photoproduction on a nucleus 
can be presented in the following form in terms of $R_g(x,Q_0^2)$ [compare to Eq.~(\ref{eq:R})],
\begin{equation}
\sigma^{\gamma A \to J/\psi A}(W_{\gamma p})=[R_{g}(x,Q_0^2)]^2 \sigma^{\gamma A \to J/\psi A}_{\rm IA}(W_{\gamma p}) \,,
\label{eq:R_Rg}
\end{equation}
where $\sigma^{\gamma A \to J/\psi A}_{\rm IA}(W_{\gamma p})$ is the IA cross section, see Eq.~(\ref{eq:cs_IA}), $x=M_{J/\psi}^2/W_{\gamma p}^2$, 
and $Q_0={\cal O}(m_c)$, where $m_c$ is the charm quark mass. In our analysis, we use 
$Q_0^2=3$ GeV$^2$, which was determined by requiring that $\sigma^{\gamma+p \to J/\psi+p}(W_{\gamma p}) \propto (xg_p(x,Q_0^2))^2$, where 
$g_p(x,Q_0^2)$ is the LO gluon density of the proton, correctly reproduces the $W_{\gamma p}$ dependence of this cross section measured at HERA~\cite{Guzey:2013qza}.
While it can be realized with a wide range of modern LO proton PDFs, 
for the calculation of $R_{g}(x,Q_0^2)$ we use the CTEQ6L1 PDFs~\cite{Pumplin:2002vw} as well as the LO H1 diffractive PDFs~\cite{H1:2006zyl} of the proton.
Note that our procedure is based on the data-driven parametrization of $d\sigma^{\gamma+p \to J/\psi+p}(W_{\gamma p},t=0)/dt$   
such that the factorization scale $Q_0$ enters only through the ratio of the nuclear and proton distributions $R_g(x,Q_0^2)$.
The variation of $Q_0$ leads to reasonably small variations of $R_g(x,Q_0^2)$, which are compatible to the uncertainty due to the 
variation of $\sigma_{\rm soft}(x,Q_0^2)$, see the discussion below.

Note that in Eq.~(\ref{eq:R_Rg}) we neglected a possible correction factor $\kappa_{A/N} \approx 1$~\cite{Guzey:2013qza}, which phenomenologically takes 
into account a deviation of the usual nuclear gluon distribution from the generalized one, which is commonly modeled using the so-called Shuvaev transform~\cite{Shuvaev:1999ce,Martin:2007sb}. The analysis in~\cite{Eskola:2023oos} has demonstrated that the two distributions are in practice 
very close at small $x$ and $\xi$ ($\xi$ is the skewness momentum fraction). Hence, with a few percent accuracy, $\kappa_{A/N}=1$.

\begin{figure}[t]
\centerline{%
\includegraphics[width=9cm]{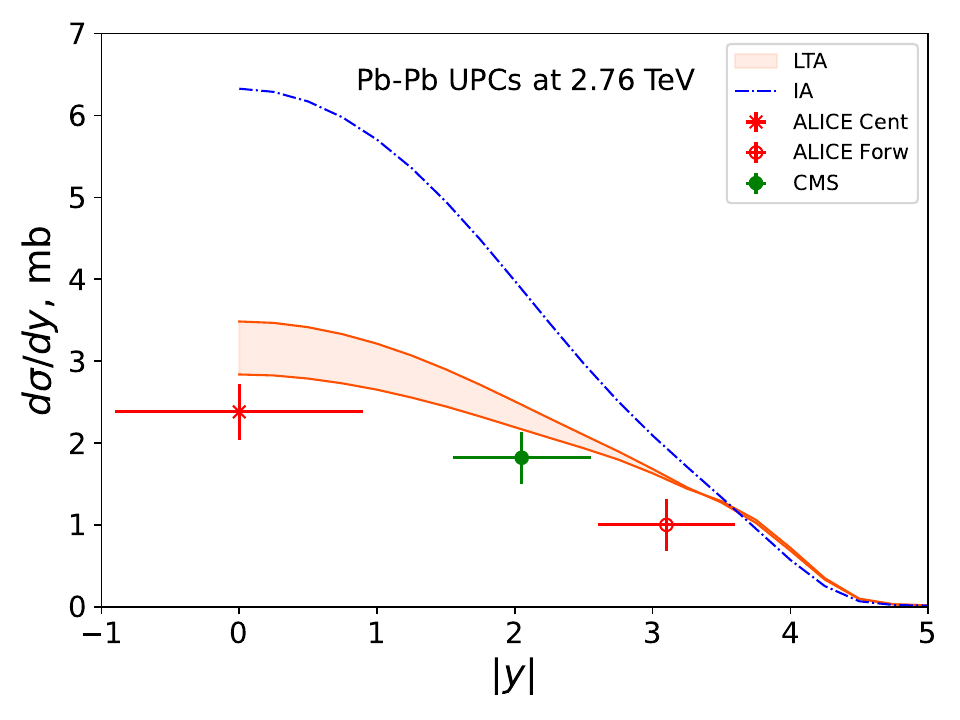}
\includegraphics[width=9cm]{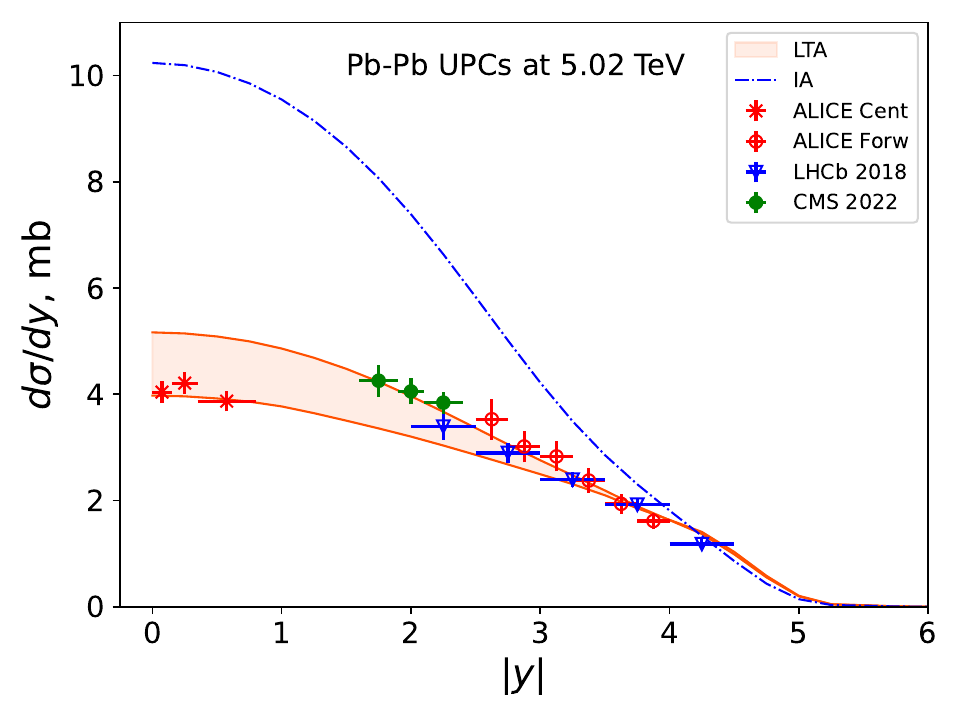}
}
\caption{The $d\sigma^{AA \to J/\psi AA}/dy$ cross section of coherent $J/\psi$ photoproduction in Pb-Pb UPCs at 2.76 TeV (left panel) and
5.02 TeV (right panel). The LTA predictions given by the shaded band are compared 
with the available Run 1~\cite{ALICE:2013wjo,ALICE:2012yye,CMS:2016itn} and Run 2~\cite{ALICE:2021gpt,ALICE:2019tqa,LHCb:2022ahs,CMS:2023snh} data. The blue dot-dashed curve is the IA result, where the gluon nuclear shadowing is turned off.
}
\label{fig:ydep_lta}
\end{figure}

Figure~\ref{fig:ydep_lta} presents the LTA predictions for the cross section of coherent $J/\psi$ photoproduction in Pb-Pb UPCs as a function
of the $J/\psi$ rapidity $|y|$ at $\sqrt{s_{NN}}=2.76$ TeV (left panel) and $\sqrt{s_{NN}}=5.02$ TeV (right panel), 
which are obtained by substituting Eq.~(\ref{eq:R_Rg}) in Eq.~(\ref{eq:upc_cs}).
 The shaded bands quantify the theoretical uncertainty of the LTA predictions
for the gluon nuclear shadowing and span the range between ``high shadowing'' (lower boundary) and ``low shadowing'' (higher boundary),
which are determined by the variation of $\sigma_{\rm soft}(x,Q_0^2)$ in Eq.~(\ref{eq:g_A_lta_2}).
To illustrate the 
importance and magnitude of the gluon nuclear shadowing, we also show by the blue dot-dashed line the IA result obtained using the GKSZ parametrization of the 
$d\sigma^{\gamma p \to J/\psi p}(t=0)/dt$ cross section, see Eq.~(\ref{eq:fit_K}), where the effect of gluon shadowing is absent.
These theoretical predictions are compared 
with the available Run 1~\cite{ALICE:2013wjo,ALICE:2012yye,CMS:2016itn} (left) and Run 2~\cite{ALICE:2021gpt,ALICE:2019tqa,LHCb:2022ahs,CMS:2023snh} (right) data. 

\begin{figure}[t]
\centerline{%
\includegraphics[width=9cm]{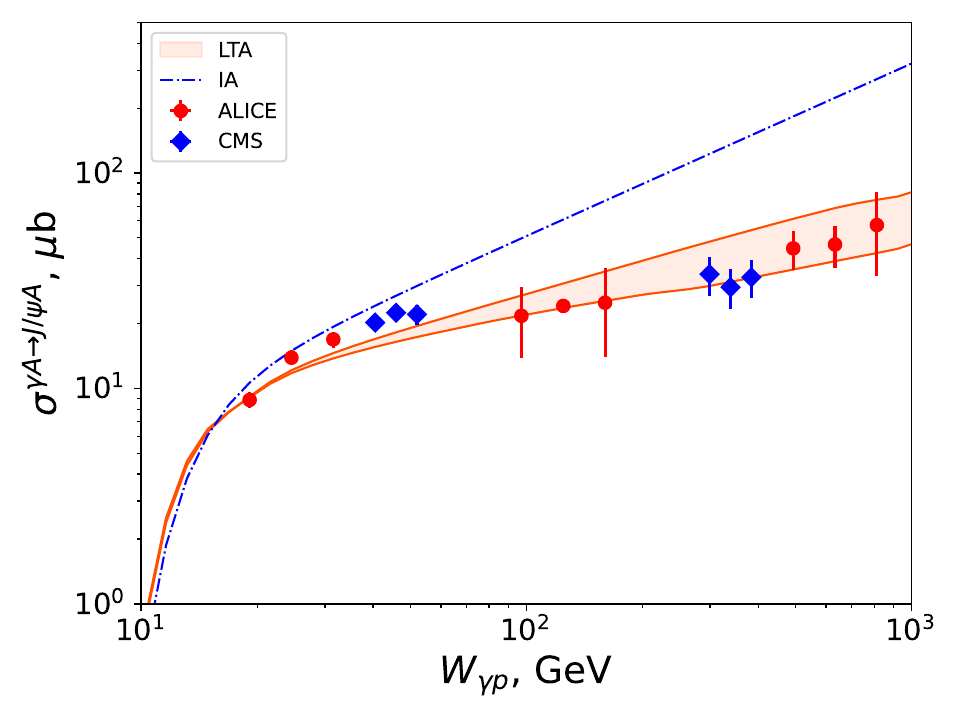}
\includegraphics[width=9cm]{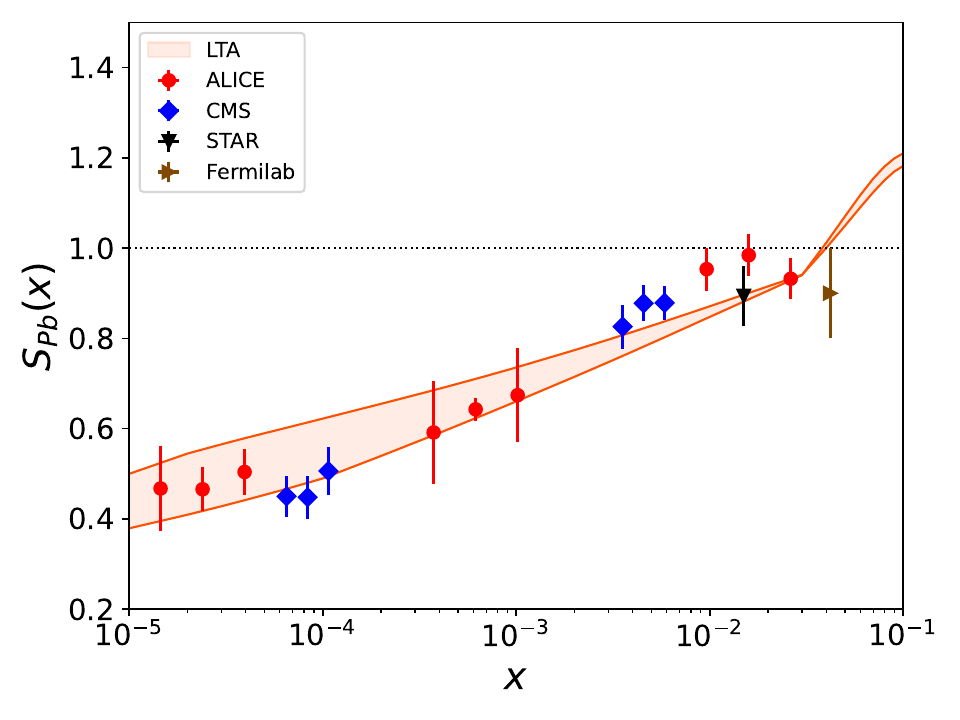}}
\caption{(Left) The nuclear photoproduction cross section $\sigma^{\gamma A \to J/\psi A}(W_{\gamma p})$
as a function of $W_{\gamma p}$: the LTA predictions (shaded band) and IA result (blue dot-dashed line) vs.~the CMS~\cite{CMS:2023snh} and ALICE~\cite{ALICE:2023jgu} data. (Right) The nuclear suppression factor $S_{Pb}(x)$ as a function of $x$: the LTA predictions (shaded band) vs.~the CMS~\cite{CMS:2023snh}, ALICE~\cite{ALICE:2023jgu}, STAR~\cite{STAR:2023gpk}, and Fermilab data~\cite{Guzey:2020ntc} [see Eq.~(\ref{eq:R_Fermilab})].
}
\label{fig:Wdep_lta}
\end{figure}

One can see from the right panel of Fig.~\ref{fig:ydep_lta} that the LTA predictions describe well both the magnitude and the rapidity dependence
of all Run 2 LHC data on $d\sigma^{AA \to J/\psi AA}/dy$. The dramatic difference between the LTA and IA results gives a clear indication
of the large gluon nuclear shadowing effect. 

At the same time, while LTA describes well the rapidity $y$ dependence of the Run 1 LHC data in the left panel of Fig.~\ref{fig:ydep_lta}, 
the predicted magnitude of $d\sigma^{AA \to J/\psi AA}/dy$ is somewhat higher than that in~\cite{Guzey:2013qza} obtained using the same approach. A detailed examination shows that this is a result of setting $\kappa_{A/N}=1$ in the present analysis (see the discussion of Eq.~(\ref{eq:R_Rg}) above) and the use of different parametrizations of $d\sigma^{\gamma p \to J/\psi p}(t=0)/dt$ for small $W_{\gamma p}$ corresponding to large $|y|$, see the discussion in Sec.~\ref{subsec:fit}.

Note that for $|y| > 3$ in the left panel and for $|y| > 4$ in the left panel, the LTA predictions lie slightly above the IA curves: this is the effect of the gluon antishadowing, see Eq.~(\ref{eq:anti}).

The left panel of Fig.~\ref{fig:Wdep_lta} presents the LTA predictions for the nuclear photoproduction cross section $\sigma^{\gamma A \to J/\psi A}(W_{\gamma p})$
as a function of $W_{\gamma p}$ and compares them with the CMS~\cite{CMS:2023snh} (blue solid diamonds) and ALICE~\cite{ALICE:2023jgu} (red solid circles) data. 
To illustrate the magnitude of nuclear modifications of $\sigma^{\gamma A \to J/\psi A}(W_{\gamma p})$, we show the IA for this cross section by the blue 
dot-dashed line. One can see from the this figure that LTA provides a good description of the data for $W_{\gamma p} > 100$ GeV, which underlines 
the importance of large leading twist gluon nuclear shadowing. At the same time, for $W_{\gamma p} < 52$ GeV, LTA underestimates $\sigma^{\gamma A \to J/\psi A}(W_{\gamma p})$ because of its appreciable gluon shadowing around $x \approx 10^{-2}$. For this range of $W_{\gamma p}$, the CMS and ALICE data (except for the 
$W_{\gamma p}=19$ GeV ALICE point) lie between the LTA and IA curves.

The experimental values for $\sigma^{\gamma A \to J/\psi A}(W_{\gamma p})$ can be converted into the nuclear suppression factor $S_{Pb}(x)$, see Eq.~(\ref{eq:R}).
The right panel of Fig.~\ref{fig:Wdep_lta} shows the LTA predictions for the gluon modification factor $R_g(x,Q_0^2)$ as a function of $x$ and its comparison
with the CMS~\cite{CMS:2023snh}, ALICE~\cite{ALICE:2023jgu}, STAR~\cite{STAR:2023gpk}, and Fermilab data~\cite{Guzey:2020ntc} [see Eq.~(\ref{eq:R_Fermilab})].
One can see from the figure that LTA describes well the $x$-dependence and magnitude of $S_{Pb}(x)$ in the wide interval of $10^{-5} < x < 0.03$. 
One should emphasize that LTA predictions have smallest theoretical uncertainties for $x \sim 10^{-3}$ because experimental uncertainties in the diffractive PDFs are small,
the model-dependent contribution due to interactions with $N \geq 3$ nucleons is a correction, and the gluon antishadowing does not play a role.
The description
is somewhat worse around $x \approx 10^{-2}$, where the data fluctuate, and the predicted nuclear suppression of $R_g(x,Q_0^2)$  appears to be 
larger than that seen in the data.

Note also that a comparison of the LTA predictions with the Fermilab data point should be taken with caution since 
possible higher-twist effects may be not negligible in this kinematics.

\subsection{Modified LTA predictions: Dynamical model for gluon antishadowing and impact parameter dependent nuclear shadowing}
\label{subsec:lta_mod}

As an alternative to the model constraining gluon antishadowing using the momentum sum rule for nuclear PDFs, see the discussion in Sec.~\ref{subsec:lta_theory}, a dynamical model of antishadowing was suggested in~\cite{Frankfurt:2016qca}. It is based on the observation 
that the physical mechanisms of both nuclear shadowing and antishadowing are based on merging of two parton ladders belonging to two different
nucleons in the target nucleus, which are close in the rapidity space. 
In practical terms, it means that the gluon antishadowing has a wider $x$-support than that in Eq.~(\ref{eq:anti}), which leads to its earlier onset at lower $x$ and, correspondingly, to somewhat smaller nuclear shadowing in the $x > 10^{-2}$ region.
The result of~\cite{Frankfurt:2016qca} can be approximated by the following numerical parametrization for $R_g^{\rm anti}(x,Q_0^2)$ in the 
$0.005 < x < 0.2$ interval [compare to Eq.~(\ref{eq:anti})],
\begin{equation}
R_g^{\rm anti}(x,Q_0^2)= N_{\rm anti}\frac{(x-0.005) (0.2-x)^2}{0.2-0.005} \,,
\label{eq:anti_2}
\end{equation}
where $N_{\rm anti}=25-29$. These values are very close to those resulting from Eq.~(\ref{eq:anti}), which indicates that the momentum sum rule of nuclear PDFs is not very sensitive to the shape of the gluon antishadowing.

Equation~(\ref{eq:R_Rg}) and its ensuing application assume that the $t$-dependence of the nuclear gluon distribution is given by 
the nuclear form factor $F_A(t)$, which is a good approximation in the case of weak nuclear shadowing. However, when the effect of shadowing is large, one needs to include its dependence on the impact parameter $\vec{b}$ (on the nucleon position in the transverse plane)~\cite{Guzey:2016qwo}.
It amounts to introducing the correction factor $[R_{g}^{\prime}(x,Q_0^2)]^2$ in the right-hand side of Eq.~(\ref{eq:R_Rg}), 
\begin{equation}
[R_{Pb}^{\prime}(x,Q_0^2)]^2=\frac{\int_{|t_{\rm min}|}^{\infty} dt\, [g_A(x,t,Q_0^2)]^2}{ [g_A(x,Q_0^2)]^2 \int_{|t_{\rm min}|}^{\infty} dt\,|F_A(t)|^2} \,,
\label{eq:R_g_prime}
\end{equation}
where $g_A(x,t,Q_0^2)$ is the nucleus generalized parton distribution (GPD) in the limit, when both gluon lines carry the same momentum fraction $x$, i.e.,  
the skewness $\xi =0$ limit. In this special case, the nuclear gluon GPD can be expressed through the impact parameter nuclear gluon distribution 
$g_A(x,\vec{b},Q_0^2)$~\cite{Frankfurt:2011cs}
[compare to Eq.~(\ref{eq:g_A_lta_2})],
\begin{equation}
g_A(x,\vec{b},Q_0^2)=g_p(x,Q_0^2)\left[\left(1-\frac{\sigma_2(x,Q_0^2)}{\sigma_{\rm soft}(x,Q_0^2)}\right) T_A(\vec{b})+2 \frac{\sigma_2(x,Q_0^2)}{\sigma_{\rm soft}^2(x,Q_0^2)} \Re e 
\left(1-e^{-\frac{1-i \eta}{2} \sigma_{\rm soft}(x,Q_0^2) T_A(\vec{b})} \right) \right]\,.
\label{eq:g_A_b}
\end{equation}
It represents the distribution of gluons in a nucleus with the momentum fraction $x$ at the transverse distance $\vec{b}$ from its center-of-mass
and is related to the nuclear gluon GPD through $g_A(x,t,Q_0^2)=\int d^2 \vec{b} \, e^{i \vec{q}_{\perp} \cdot \vec{b}} g_A(x,\vec{b},Q_0^2)$.

Nuclear shadowing is stronger at the center of the nucleus ($|\vec{b}| \approx 0$) than at its periphery ($|\vec{b}| \approx R_A$), which brings non-trivial
correlations between $x$ and $\vec{b}$ encoded in the $R_{Pb}^{\prime}(x,Q_0^2)$ factor. In particular, it leads to a broadening of the nuclear gluon distribution in the impact parameter space, which manifests itself in the shift of the diffractive minima of the differential
cross section of coherent $J/\psi$ photoproduction on a nucleus, $d\sigma^{\gamma A \to J/\psi A}(W_{\gamma p},t)/dt$,
toward smaller values of $|t|$~\cite{Guzey:2016qwo}. This prediction has been confirmed by the ALICE measurement~\cite{ALICE:2021tyx}.

At the same time, the role of the discussed effect in the $t$-integrated $\sigma^{\gamma A \to J/\psi A}(W_{\gamma p})$ cross section used in our analysis is modest. The factor $[R_{Pb}^{\prime}(x,Q_0^2)]^2]$ suppresses $\sigma^{\gamma A \to J/\psi A}(W_{\gamma p})$ by $6-8$\%
for $10^{-5} < x < 10^{-2}$ and then $R_{Pb}^{\prime}(x,Q_0^2) \to 1$, when $x \to x_0=0.1$.

\begin{figure}[t]
\centerline{%
\includegraphics[width=9cm]{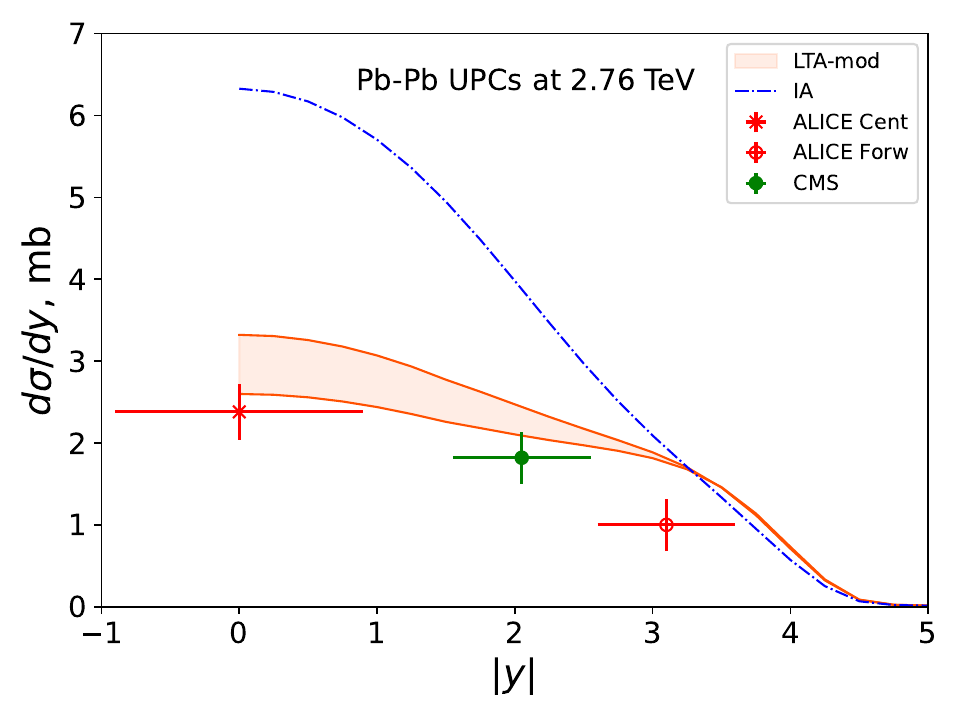}
\includegraphics[width=9cm]{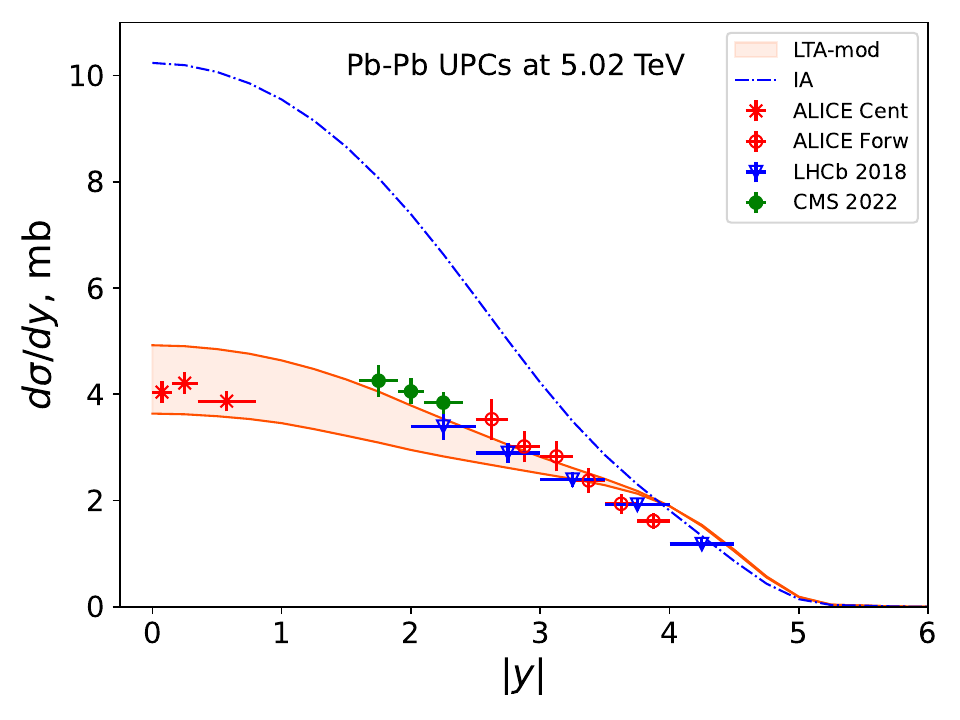}
}
\centerline{%
\includegraphics[width=9cm]{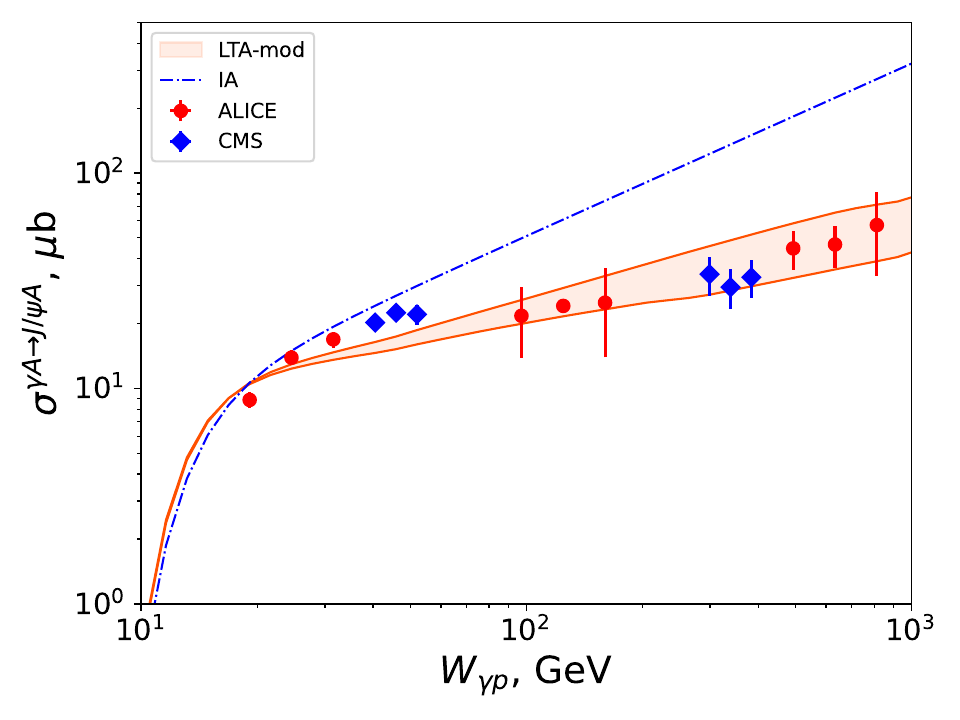}
\includegraphics[width=9cm]{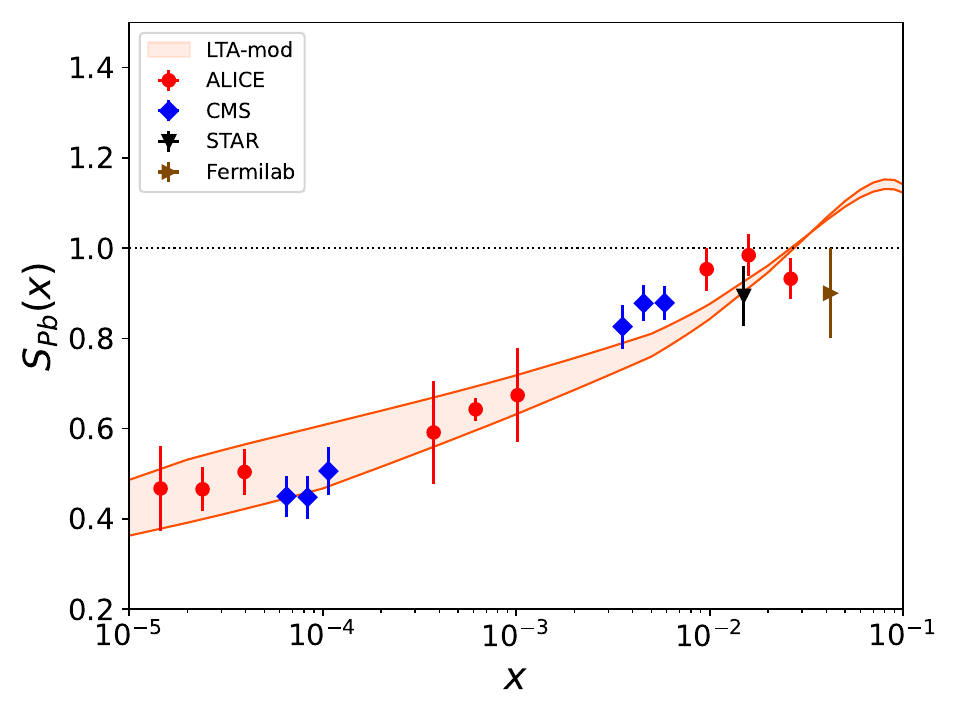}}
\caption{Modified LTA predictions for $d\sigma^{AA \to J/\psi AA}/dy$ as a function of $|y|$ at 2.76 TeV (upper left) and
5.02 TeV (upper right), $\sigma^{\gamma A \to J/\psi A}(W_{\gamma p})$ as a function of $W_{\gamma p}$ (lower left), and 
$S_{Pb}(x)$ as a function of $x$ (lower right) and their comparison with the available UPC data, see Figs.~\ref{fig:ydep_lta} and
\ref{fig:Wdep_lta} for references. The blue dot-dashed line is the IA result neglecting gluon nuclear shadowing.
}
\label{fig:ydep_lta_mod}
\end{figure}

Figure~\ref{fig:ydep_lta_mod} presents the modified LTA predictions for the UPC cross section $d\sigma^{AA \to J/\psi AA}/dy$ as a function of $|y|$ at 2.76 TeV (upper left) and 5.02 TeV (upper right), the photoproduction cross section $\sigma^{\gamma A \to J/\psi A}(W_{\gamma p})$ as a function of $W_{\gamma p}$ (lower left), and  the nuclear suppression factor $S_{Pb}(x)$ as a function of $x$ (lower right). The shaded bands quantify the theoretical uncertainty of the LTA predictions, with the upper and lower boundaries corresponding to the ``low shadowing'' and ``high shadowing'' cases, respectively.
The blue dot-dashed line is the IA result, where one neglects the effect of the leading twist gluon nuclear shadowing. 
These theoretical results are compared to the available UPC data for these observables, see Figs.~\ref{fig:ydep_lta} and
\ref{fig:Wdep_lta} for references.  

Compared to the LTA results shown in Figs.~\ref{fig:ydep_lta} and
\ref{fig:Wdep_lta}, the modified LTA predictions differ in two aspects. First, an earlier onset of the gluon antishadowing modeled using Eq.~(\ref{eq:anti_2})
somewhat worsens the agreement with the data on $d\sigma^{AA \to J/\psi AA}/dy$ at $|y| \approx 3$ (Run 1) and $|y| \approx 4$ (Run 2) and simultaneously improves the agreement with the data on $\sigma^{\gamma A \to J/\psi A}(W_{\gamma p})$ at low $W_{\gamma p}$ and $S_{Pb}(x)$ at $x \approx 10^{-2}$.
Second, the correction for the impact parameter dependent nuclear shadowing, see Eq.~(\ref{eq:R_g_prime}), slightly lowers $d\sigma^{AA \to J/\psi AA}/dy$ 
at central rapidities, $\sigma^{\gamma A \to J/\psi A}(W_{\gamma p})$ at large $W_{\gamma p}$ and $S_{Pb}(x)$ at small $x$ and, hence, makes the agreement 
of the LTA predictions with the UPC data even more convincing.

\subsection{Nuclear suppression factor from global QCD analyses of nuclear PDFs}
\label{subsec:global_fits}

It is important to compare the nuclear suppression factor $S_{Pb}(x)$ with patterns of nuclear modifications of the nuclear gluon 
distribution at small $x$, which emerge from global QCD analyses of a wide array of fixed-target and collider data~\cite{Eskola:2021nhw,Kovarik:2015cma,AbdulKhalek:2022fyi}, for a review, see~\cite{Klasen:2023uqj}.

\begin{figure}[t]
\centerline{%
\includegraphics[width=11cm]{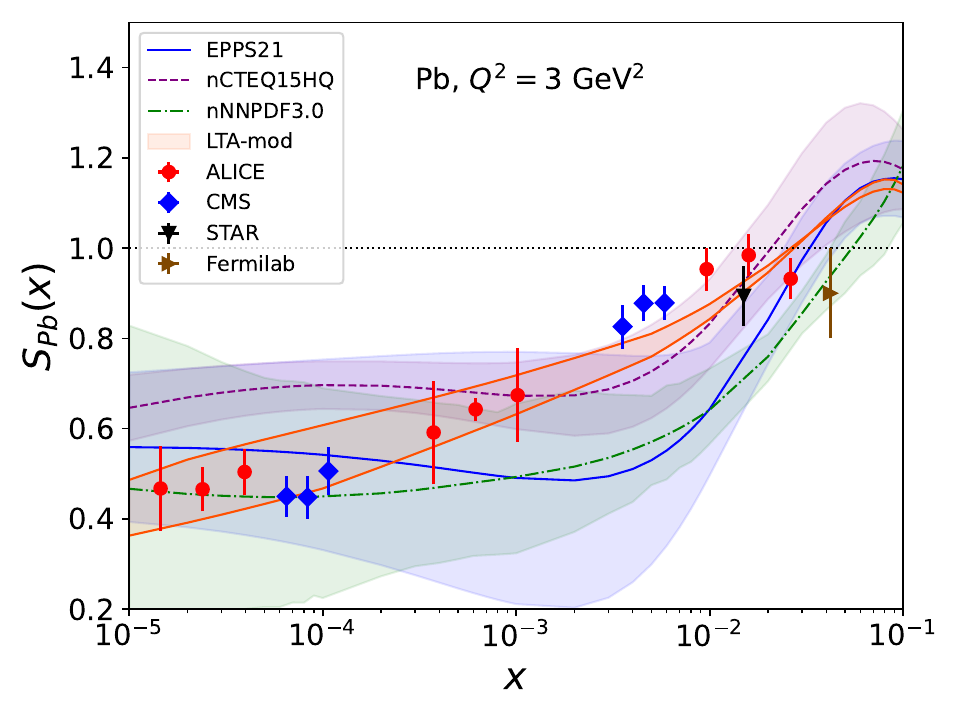}}
\caption{The EPPS21, nCTEQ15HQ, nNNPDF3.0, and modified  LTA results for $R_g(x,Q_0^2)$ as a function of $x$ at $Q_0^2=3$ GeV$^2$ for $^{208}$Pb and their comparison to the nuclear suppression factor $S_{Pb}(x)$, see Fig.~\ref{fig:Wdep_lta} for details.
}
\label{fig:ydep_lta_mod3}
\end{figure}

Figure~\ref{fig:ydep_lta_mod3} compiles predictions of three modern, state-of-the-art analyses of nuclear PDFs, EPPS21~\cite{Eskola:2021nhw},
nCTEQ15HQ~\cite{Kovarik:2015cma,Duwentaster:2022kpv}, and nNNPDF3.0~\cite{AbdulKhalek:2022fyi}, for the ratio
of next-to-leading order (NLO) gluon distributions in $^{208}$Pb and the proton, $R_g(x,Q_0^2)$, as a function of $x$ at $Q^2_0=3$ GeV$^2$.
The shaded bands represent the uncertainties of the nuclear PDF calculated using corresponding error PDFs.
They are compared with the modified LTA predictions for $R_g(x,Q_0^2)$ discussed in Sec.~\ref{subsec:lta_mod} and the experimental values for the nuclear suppression factor $S_{Pb}(x)$, see Fig.~\ref{fig:Wdep_lta}. We discussed in Sec.~\ref{subsec:lta_comp} that the NLO
corrections for exclusive $J/\psi$ photoproduction are very large, which challenges the interpretation of these data in terms of the gluon distribution
of the target. Nevertheless, it is instructive to compare the predictions for $R_g(x,Q_0^2)$, which have not used UPC data,
with $S_{Pb}(x)$ emerging from the 
data on coherent $J/\psi$ photoproduction in heavy-ion UPCs.

One can see from the figure that within large uncertainties of nuclear PDFs (the uncertainties are especially large because $Q^2_0=3$ GeV$^2$ is close to the initial
scale of these PDFs), the EPPS21 and nNNPDF3.0
nuclear PDFs are consistent with $S_{Pb}(x)$ extracted from the UPC data as well as with the LTA result in the entire range of $x$.
The nCTEQ15HQ prediction, which corresponds to a weaker gluon shadowing, lies above the data for $x< 10^{-4}$.
Note that, by construction, 
the EPPS21, nCTEQ15HQ, and nNNPDF3.0 parametrizations predict the almost constant value of $R_g(x,Q_0^2)$ for $x < 5 \times 10^{-3}$, while the data and 
the $\chi^2$ fit to them clearly favor a decreasing $S_{Pb}(x)$ as $x$ decreases, see Sec.~\ref{subsec:fit}.
Nuclear suppression also increases with a decrease of $x$ in the LTA framework, where it is explained by an increasing relative probability of 
hard diffraction encoded in the the $\sigma_2(x,Q_0^2)$ cross section, see Eq.~(\ref{eq:sigma_2}).

\section{Conclusions}
\label{sec:conclusions}

Photoproduction of light and heavy vector mesons on nuclei is one of central processes in the program of UPC measurements at the LHC and RHIC.
The experimental observation that the cross section of coherent $J/\psi$ photoproduction in heavy-ion UPCs is significantly suppressed compared to the 
impulse approximation expectation provides an important input for theoretical studies of QCD at small $x$ and challenges both the leading twist and gluon saturation pictures of this process. Hence, for an unambiguous interpretation of these data, it is important to determine the nuclear suppression factor 
$S_{Pb}(x)$, which quantifies nuclear modifications of this cross section, in a model-independent way.

In this paper, we determine $S_{Pb}(x)$ by performing the $\chi^2$ fit to all available data on the UPC cross section $d\sigma^{AA \to J/\psi AA}/dy$ 
as a function of the $J/\psi$ rapidity $|y|$ and the photoproduction cross section $\sigma^{\gamma A \to J/\psi A}(W_{\gamma p})$ as a function of the photon-nucleon energy $W_{\gamma p}$. We find that while the $d\sigma^{AA \to J/\psi AA}/dy$  data alone constrain $S_{Pb}(x)$ for $x \geq 10^{-3}$,
the combined $d\sigma^{AA \to J/\psi AA}/dy$  and $\sigma^{\gamma A \to J/\psi A}(W_{\gamma p})$ data allow us to determine
$S_{Pb}(x)$ in the wide interval $10^{-5} < x < 0.05$. In particular, the data favor $S_{Pb}(x)$, which decreases with a decrease of $x$ in the $10^{-4} < x < 0.01$ 
interval. For $ x < 10^{-4}$, both constant and decreasing $S_{Pb}(x)$ can be accommodated by the data.
We illustrate how predictions based on different fit solutions describe $d\sigma^{AA \to J/\psi AA}/dy$ and $\sigma^{\gamma A \to J/\psi A}(W_{\gamma p})$ used in the fit as well the experimental values of $S_{Pb}(x)$ derived from $\sigma^{\gamma A \to J/\psi A}(W_{\gamma p})$.

We identify $S_{Pb}(x)$ with $R_g(x,Q_0^2)=g_A(x,Q_0^2)/[A g_p(x,Q_0^2)]$, where $g_A(x,Q_0^2)$ and  $g_p(x,Q_0^2)$ are the gluon distributions in a nucleus and the proton, respectively, and $Q_0^2=3$ GeV$^2$ is the resolution scale set by the charm quark mass, which allows us to interpret the UPC data in terms of nuclear PDFs. We find that the leading twist approximation (LTA) for nuclear shadowing provides a good description of all the data on $d\sigma^{AA \to J/\psi AA}/dy$, $\sigma^{\gamma A \to J/\psi A}(W_{\gamma p})$ and $S_{Pb}(x)$, which
further supports the evidence of strong leading-twist gluon nuclear shadowing at small $x$~\cite{Guzey:2013xba,Guzey:2013qza}. 
One should emphasize that the LTA description of $S_{Pb}(x)$ at intermediate $x \approx 10^{-2}$ is also fair and demonstrates that it can be further improved by dynamical modeling of the gluon antishadowing. 
In the opposite small-$x$ limit, the small correction for the impact parameter dependence of nuclear shadowing makes the agreement of the LTA predictions with the experimental values for $S_{Pb}(x)$ even more convincing.

Finally, we also show that within large uncertainties of modern state-of-the-art nuclear PDFs, the EPPS21 and nNNPDF3.0 nuclear PDFs give a reasonable description of $S_{Pb}(x)$ in the entire range of $x$; the nCTEQ15HQ PDFs with a weaker gluon shadowing somewhat overestimate the data for $x < 10^{-4}$.

Note that we used the ALICE data on the $t$ dependence of coherent $J/\psi$ photonuclear production~\cite{ALICE:2021tyx} only indirectly, 
as an argument supporting introduction of the $R_{Pb}^{\prime}(x,Q_0^2)$ correction factor, see Eq.~(\ref{eq:R_g_prime}). The full use of these data 
requires a separate dedicated analysis, where the fitting function contains additional free parameters modeling the $\vec{b}$ dependence of the nuclear suppression factor. 

As we explained in Introduction, predictions of the framework based on the collinear QCD factorization and leading twist nuclear PDFs 
for coherent $J/\psi$ photoproduction in heavy-ion UPCs can be contrasted with those based on the color dipole model. A summary of results of the color glass condensate approach for the UPC observables considered in this work is presented in~\cite{Mantysaari:2023xcu}.

\acknowledgments

The research of V.G.~was funded by the Academy of Finland project 330448, the Center of Excellence in Quark Matter
of the Academy of Finland (projects 346325 and 346326), and the European Research Council project ERC-2018-ADG-835105
YoctoLHC. The research of M.S.~was supported by the US Department of Energy Office
of Science, Office of Nuclear Physics under Award No. DE- FG02-93ER40771.

\end{document}